\documentclass[sigconf]{acmart}
\pdfoutput=1

\setcopyright{rightsretained}

\acmDOI{10.475/123_4}

\acmISBN{123-4567-24-567/08/06}

\renewcommand\footnotetextcopyrightpermission[1]{} 

\usepackage{amsmath,amssymb}
\usepackage{array}
\usepackage{mdwtab}
\usepackage[caption=false,font=footnotesize]{subfig}
\usepackage{fixltx2e}
\usepackage{dblfloatfix} 

\usepackage{xcolor}
\usepackage{xspace}
\usepackage{numprint}
\usepackage{algorithm}
\usepackage{algorithmicx}
\usepackage{algpseudocode}
\usepackage{float}
\usepackage{placeins}
\newfloat{algorithm}{t}{lop}
\usepackage{graphicx}
\usepackage{subfig}

\usepackage{flushend}

\newcommand{\e}{\mathbf{e}}

\newcommand{\NP}{$\mathcal{NP}$}

\newcommand{\bigO}{\mathcal{O}}

\newcommand{\iec}{\textit{i.\,e.},\xspace}
\newcommand{\ie}{\textit{i.\,e.}\xspace}

\newcommand{\eg}{\textit{e.\,g.}\xspace}

\newcommand{\etal}{\textit{et al.}\xspace}

\newcommand{\wrt}{w.\,r.\,t.\xspace}

\newcommand{\mswap}{\textsc{TiMEr}\xspace}

\newcommand{\libtopomap}{\textsc{LibTopoMap}\xspace}

\newcommand{\kahip}{\textsc{KaHIP}\xspace}

\newcommand{\scotch}{\textsc{Scotch}\xspace}

\newcommand{\djoko}{Djokovi\'{c} relation\xspace}
\newcommand{\djokoRelated}{Djokovi\'{c}\xspace related\xspace}

\newcommand{\argmin}{\operatorname{argmin}\xspace}
\newcommand{\argmax}{\operatorname{argmax}\xspace}
\newcommand{\cc}{\operatorname{Coco}\xspace}
\newcommand{\divers}{\operatorname{Div}\xspace}
\newcommand{\ccd}{\operatorname{Coco^+}\xspace}

\newcommand{\contract}{\operatorname{contract}\xspace}
\newcommand{\assemble}{\operatorname{assemble}\xspace}
\newcommand{\modulo}{\operatorname{mod}\xspace}

\newcommand{\parent}{\operatorname{parent}\xspace}
\newcommand{\oldParent}{\operatorname{oldParent}\xspace}
\newcommand{\newParent}{\operatorname{newParent}\xspace}
\newcommand{\newParentLabel}{\operatorname{newParentLabel}\xspace}
\newcommand{\prefLabel}{\operatorname{prefLabel}\xspace}

\newcommand{\initial}{\textsc{Identity}\xspace}

\newcommand{\greedyallcM}{\textsc{GreedyAllC} \xspace}

\newcommand{\greedymin}{\textsc{GreedyMin}\xspace}

\newcommand{\cone}{\mathtt{c1}\xspace}
\newcommand{\ctwo}{\mathtt{c2}\xspace}
\newcommand{\cthree}{\mathtt{c3}\xspace}
\newcommand{\cfour}{\mathtt{c4}\xspace}

\usepackage[colorinlistoftodos,prependcaption,textsize=tiny]{todonotes}

\usepackage{times} 

\begin{document}

\title{Topology-induced Enhancement of Mappings}
\titlenote{This work is partially supported by German Research Foundation (DFG) grant ME 3619/2-1.}

\author{Roland Glantz}
\affiliation{%
  \institution{Karlsruhe Institute of Technology}
  \city{Karlsruhe}
  \country{Germany}
}
\email{rolandglantz@gmail.com}

\author{Maria Predari}
\affiliation{%
  \institution{University of Cologne}
  \city{Cologne}
  \country{Germany}
}
\email{mpredari@uni-koeln.de}

\author{Henning Meyerhenke}
\affiliation{%
  \institution{University of Cologne}
  \city{Cologne}
  \country{Germany}
}
\email{h.meyerhenke@uni-koeln.de}

\renewcommand{\shortauthors}{R. Glantz et al.}

\begin{abstract}
In this paper we propose a new method to enhance a mapping
$\mu(\cdot)$ of a parallel application's computational tasks to
the processing elements (PEs) of a parallel computer.
  The idea behind our method \mswap is to enhance such a mapping by drawing on 
  the observation that many topologies take the form of a partial cube.
  This class of graphs includes all rectangular and cubic meshes, any such torus with
  even extensions in each dimension, all hypercubes, and all trees.

  Following previous work, we represent the parallel application and the parallel computer
  by graphs $G_a = (V_a, E_a)$ and $G_p = (V_p, E_p)$.
  $G_p$ being a partial cube allows us to label its vertices, the PEs,
  by bitvectors such that the cost of exchanging one unit of
  information between two vertices $u_p$ and $v_p$ of $G_p$ amounts to
  the Hamming distance between the labels of $u_p$ and $v_p$.
  
By transferring these bitvectors from $V_p$ to $V_a$ via
  $\mu^{-1}(\cdot)$ and extending them to be unique on
  $V_a$, we can enhance $\mu(\cdot)$ by swapping labels of $V_a$
  in a new way. 
  Pairs of swapped labels are local \wrt the PEs, but
  not \wrt $G_a$.
%
Moreover, permutations of the bitvectors' entries give rise to a plethora of
hierarchies on the PEs. Through these hierarchies we turn \mswap
  into a hierarchical method for improving $\mu(\cdot)$ that is
complementary to state-of-the-art methods for computing $\mu(\cdot)$
in the first place.

In our experiments we use \mswap to enhance mappings of complex
networks onto rectangular meshes and tori with 256 and 512
nodes, as well as hypercubes with 256 nodes. It turns out that
common quality measures of mappings derived from state-of-the-art
algorithms can be improved considerably.

\end{abstract}

\keywords{multi-hierarchical mapping; parallel communication optimization;
partial cube; Hamming distance}

\maketitle


%
%
\section{Introduction}
\label{sec:intro}
%
Large-scale matrix- or graph-based applications such as numerical simulations~\cite{TrobecK15parallel} 
or massive network analytics~\cite{DBLP:conf/ipps/QueCPG15} often run
on parallel systems with distributed memory. The iterative nature of the underlying algorithms
typically requires recurring communication between the processing elements (PEs). 
To optimize the running time of such an application, the computational load should be evenly 
distributed onto the PEs, while at the same time, the communication volume between them should be low.
When mapping processes to PEs on non-uniform memory access (NUMA) systems, 
it should also be taken into account that the cost for communication operations depends
on the locations of the PEs involved.
In particular, one wants to map heavily communicating processes ``close
to each other''. 

More formally, let the application be modeled by an \emph{application graph} 
$G_a = (V_a, E_a, \omega_a)$ that needs to be distributed over the PEs.
A vertex in $V_a$ represents a computational task of the application, while
an edge $e_a = \{u_a, v_a\}$ indicates data exchanges between tasks $u_a$
and $v_a$, with $\omega_a(e_a)$ specifying the amount of data to be exchanged.
The topology of the parallel system is modeled by a pro\-ces\-sor
graph $G_p = (V_p, E_p, \omega_p)$, where the edge weight
$\omega_p(\{u_p, v_p\})$ indicates the cost of exchanging one
  data unit between the PEs $u_p$ and
  $v_p$~\cite{7161509,Glantz2015c}.
%
A \emph{balanced} distribution of processes onto PEs thus corresponds
to a \emph{mapping} $\mu: V_a \mapsto V_p$ such that, for some small
$\varepsilon \geq 0$,
\begin{equation}
\label{balance_constraint}                                                                                                                                  
\vert \mu^{-1}(v_p) \vert \leq (1+\varepsilon)\cdot \lceil \vert V_a \vert / \vert \mu(V_a) \vert
\rceil
\end{equation}
for all $v_p \in \mu(V_a)$. Therefore, $\mu(\cdot)$ induces a balanced
partition of $G_a$ with blocks $\mu^{-1}(v_p)$, $v_p \in
V_p$. Conversely, one can find a mapping $\mu: V_a \mapsto V_p$ that
fulfills Eq.~(\ref{balance_constraint}) by first partitioning $V_a$
into $\vert V_p \vert$ balanced parts~\cite{GPOverviewBook} and then
specifying a one-to-one mapping from the blocks of the partition onto
the vertices of $G_p$.
%
For an accurate modeling of communication costs, one would also have to specify
the path(s) over which two vertices communicate in $G_p$.
This is, however, system-specific. For the purpose of generic
algorithms, it is thus often abstracted away by assuming routing on
shortest paths in $G_p$
~\cite{hoefler-topomap}. In this work, we make the same assumption.


In order to steer an optimization
process for obtaining good mappings, different objective functions
have been proposed~\cite{hoefler-topomap,ManKim1991246,7161509}. 
%
One of the most commonly used~\cite{Pellegrini11static} objective functions 
is $\cc(\cdot)$ (also referred to as \emph{hop-byte} in~\cite{Yu06}), cf. Eq.~(\ref{eq:objFuncCommCost2}):
\begin{align}
\label{eq:objFuncCommCost1}
\mu^* &:= \argmin\limits_{\substack{\mu: V_a \mapsto V_p\\\mu
    \mbox{\tiny~balanced}}} \cc(\mu) \mbox{, with}\\
\label{eq:objFuncCommCost2}
\cc(\mu) &:= \sum_{\substack{\e_a \in E_a\\{\mbox{\tiny $e_a = \{u_a, v_a\}$}}}} \omega_a(e_a)~d_{G_p}
(\mu(u_a), \mu(v_a)),
\end{align}
where $d_{G_p}(\mu(u_a), \mu(v_a))$ denotes the distance between
$\mu(u_a)$ and $\mu(v_a)$ in $G_p$, \ie the number of edges on
shortest paths.
Broadly speaking, $\cc(\cdot)$ is minimised when pairs of highly
  communicating processes are placed in nearby processors.
%
%
Finding $\mu^*(\cdot)$ is \NP-hard.
Indeed, 
finding $\mu^*(\cdot)$ for
a complete graph $G_p$ amounts to graph partitioning, which is an
\NP-hard problem~\cite{Garey:1979:CIG:578533}.


\paragraph{Related work and motivation}
%
One way of looking at previous algorithmic work from a high-level
perspective (more details can be found in the overview articles by
Pellegrini~\cite{Pellegrini11static},
Bulu\c{c} \etal~\cite{Buluc2013a}, and
Aubanel~\cite{Aubanel09resource}) is to group it into two
categories. One line has tried to couple mapping with
graph partitioning. 
To this end, the objective function for
  partitioning, usually the edge cut, is replaced by an objective
  function like $\cc(\cdot)$ that considers distances $d_{G_p}(\cdot)$~\cite{DBLP:journals/fgcs/WalshawC01}. 
  To avoid recomputing these values, Walshaw and
  Cross~\cite{DBLP:journals/fgcs/WalshawC01} store them in a \emph{network cost
    matrix} (NCM). 
When our proposed method \mswap (\underline{T}opology-\underline{i}nduced
  \underline{M}apping \underline{E}nhance\underline{r}) is used to enhance a mapping, it does so without an NCM, thus avoiding its quadratic space complexity.
%

\begin{figure}
\subfloat[$G_a$]{\includegraphics[width=2.6cm]{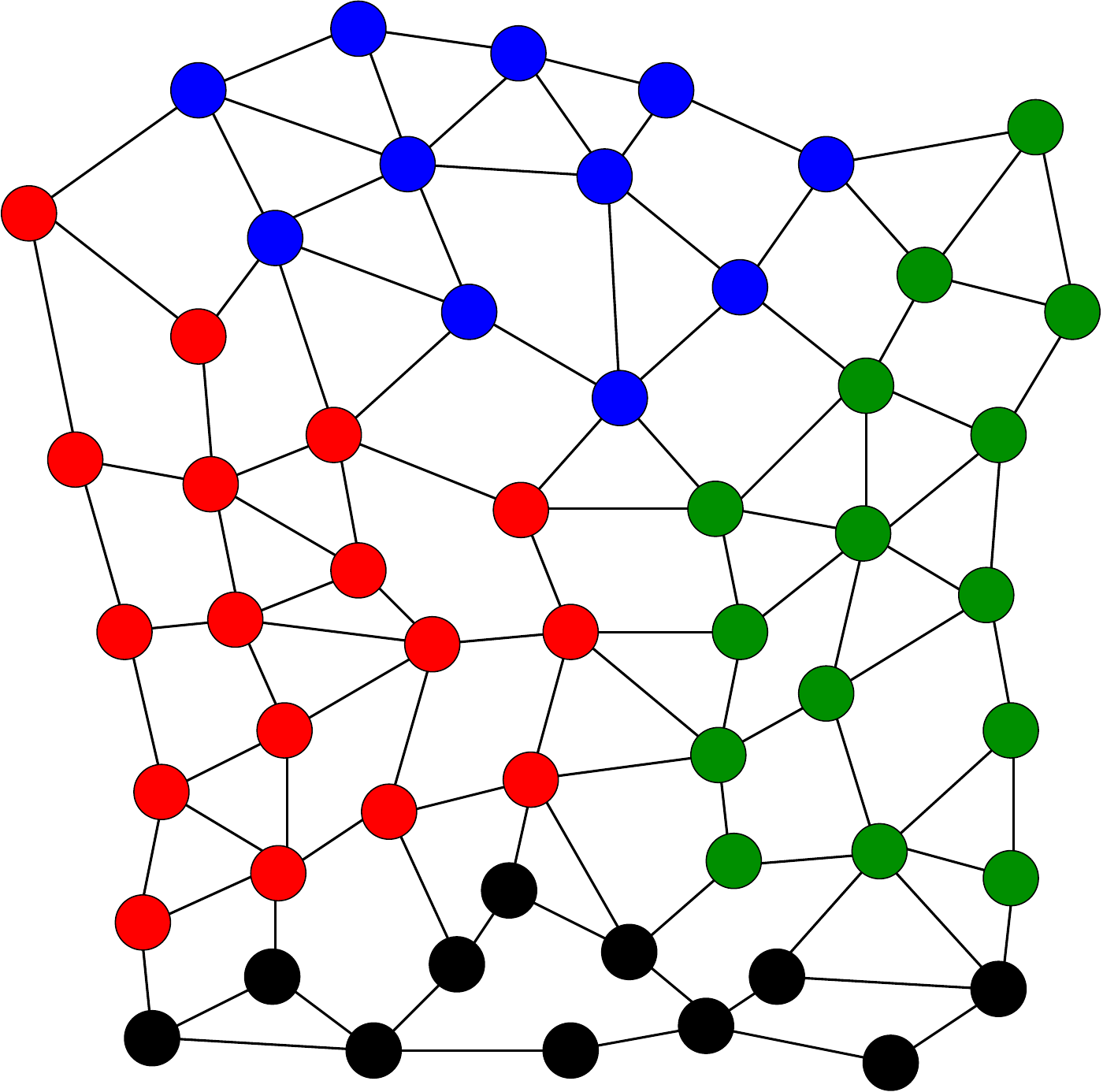}\label{1a}}
\subfloat[$G_c$]{\includegraphics[width=2.6cm]{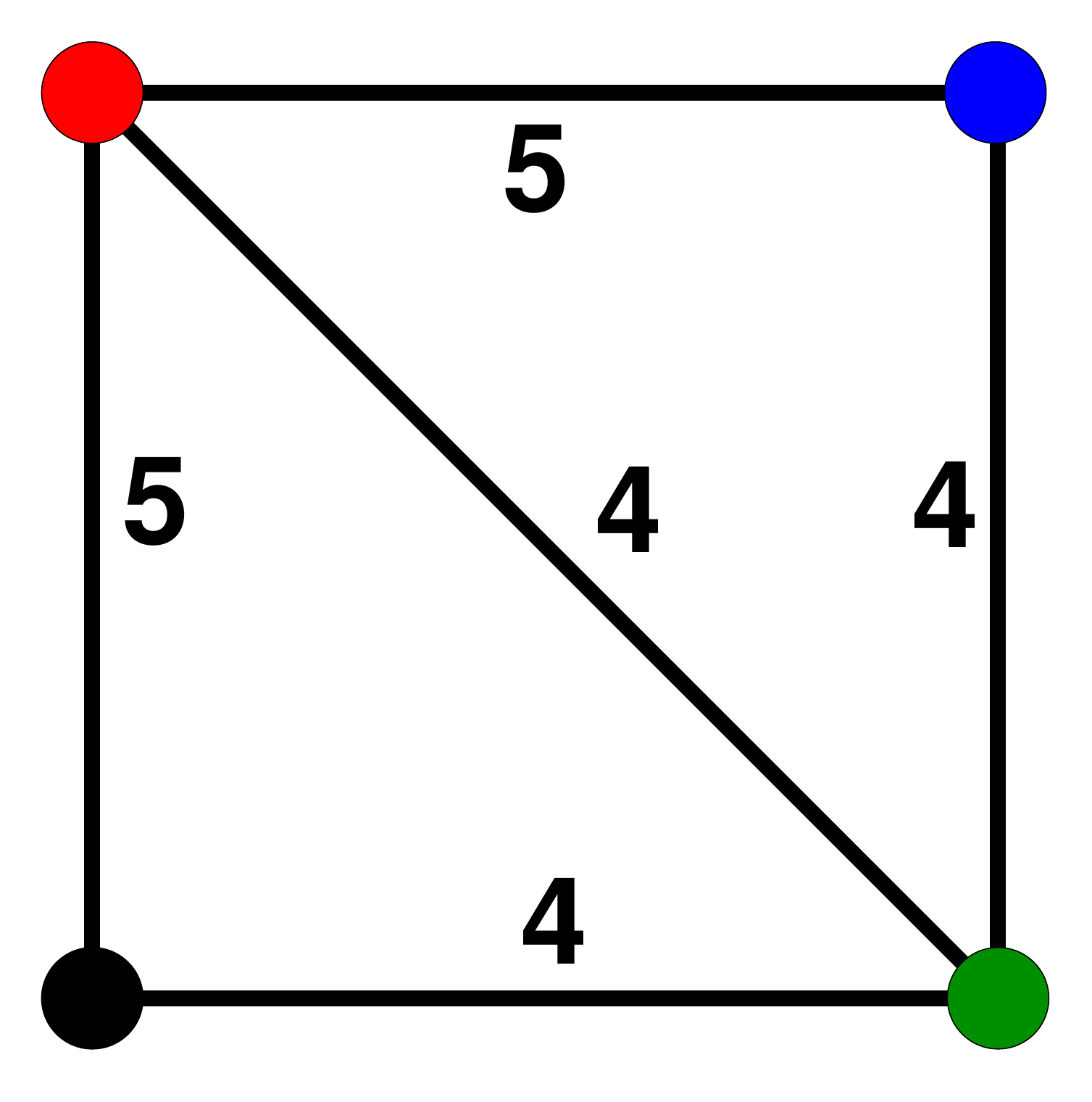}\label{1b}}
\subfloat[$G_p$]{\includegraphics[width=2.6cm]{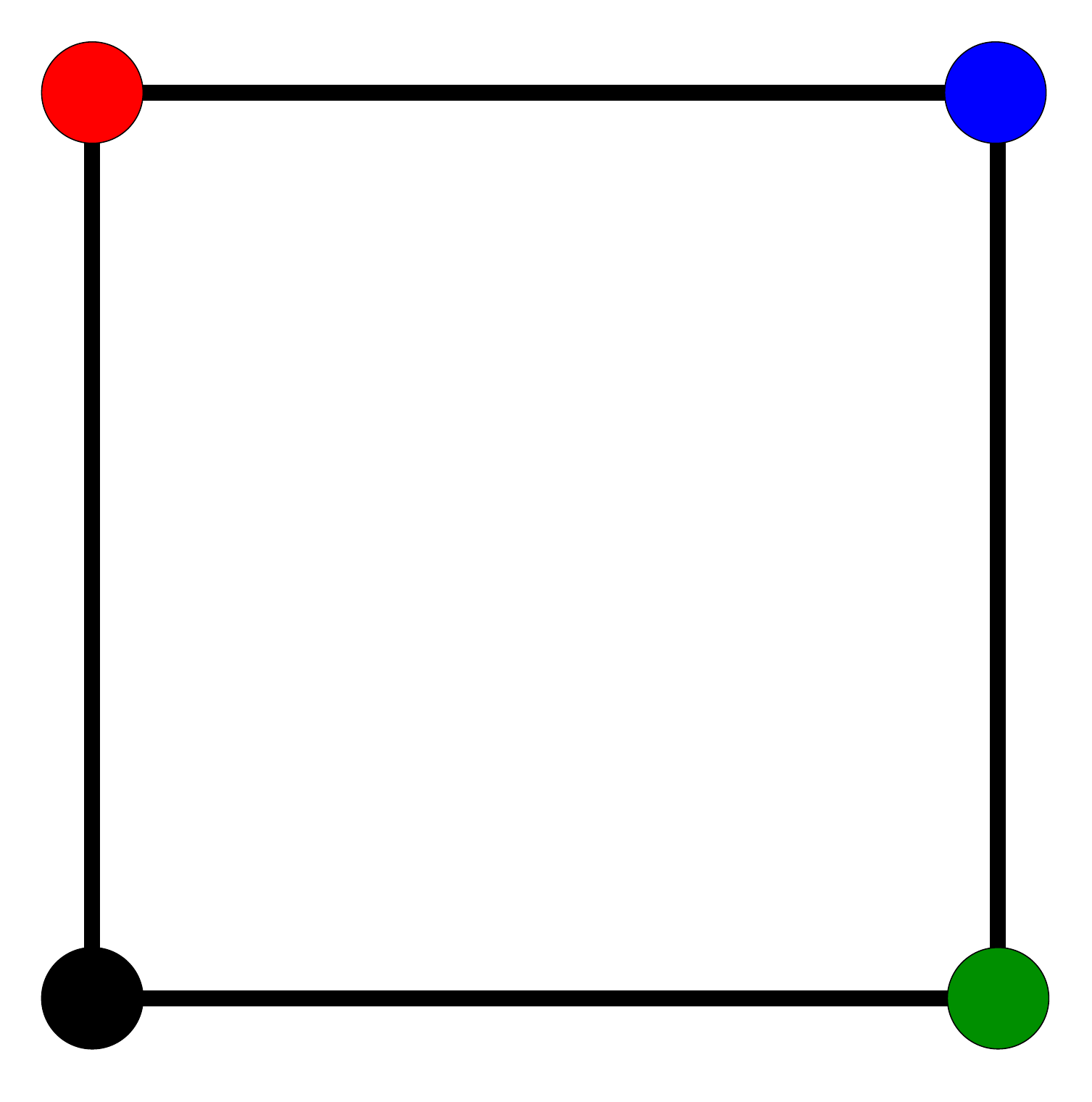} \label{1c}}   
\caption{\label{fig:mapping}
   {\small (a) Application graph partitioned into four blocks with unitary edge weights.
   (b) Resulting communication graph, numbers indicate edge weights.
   (c) The processor graph is a partial cube and the bijection $\nu(\cdot)$
    between the vertices of $G_c$ and those of $G_p$ is indicated by the colors.
    Communication between the red and the green vertex has to be routed via an intermediary
    node (the blue one or the black one).}
}
\end{figure}

The second line of research decouples partitioning and
  mapping. First, $G_a$ is partitioned into $\vert G_p \vert$ blocks
  without taking $G_p$ into account. Typically, this step involves
  multilevel approaches whose hierarchies on $G_a$ are built with 
  edge contraction~\cite{Karypis1998a,DBLP:conf/esa/OsipovS10,Sanders2013a}
  or weighted aggregation~\cite{DBLP:conf/lion/ChevalierS09,MeyerhenkeMS09new}.
  Contraction of the blocks of $G_a$ into single vertices
  yields the communication graph $G_c = (V_c, E_c, \omega_c)$,
  where $\omega_c(e_c)$, $e_c \in E_c$,
  aggregates the weight of edges in $G_a$ with end
  vertices in different blocks.
  For an example see
  Figures~\ref{1a},\ref{1b}. Finally, one computes a bijection $\nu:
  V_c \mapsto V_p$ that minimizes $\cc(\cdot)$ or a related
  function, using (for example) greedy
  methods~\cite{hoefler-topomap,Brandfass2013372,Glantz2015c,7161509}
  or metaheuristics~\cite{Brandfass2013372,Ucar200632}, see Figure~\ref{1c}. When \mswap is used to
  enhance such a mapping, it modifies not only $\nu(\cdot)$, but also
  affects the partition of $V_a$ (and thus possibly $G_c$).
  Hence, deficiencies due to the decoupling
  of partitioning and mapping can be compensated.
  Note that, since \mswap is proposed as an improvement on mappings derived from state-of-the-art methods,
  we assume that an initial mapping
is provided. This is no limitation: a mapping can be easily computed from the
solution of a graph partitioner using the identity mapping from
$G_c$ to $G_p$.

Being man-made and cost-effective, the topologies of parallel
  computers exhibit special properties that can be used to find good
  mappings. A widely used property is that PEs are often
  ``hierarchically organized into, e. g., islands, racks, nodes,
  processors, cores with corresponding communication links of similar
  quality''~\cite{Schulz2017a}. Thus, dual recursive bisection
  (DRB)~\cite{Pellegrini94static}, the recent \emph{embedded sectioning}~\cite{DBLP:journals/ijhpca/KirmaniPR17},
  and what can be called recursive multisection~\cite{chan2012impact,6495451,Schulz2017a} suggest
  themselves for mapping: DRB and embedded sectioning cut both $G_a$ (or $G_c$) and $G_p$ into two
  blocks recursively and assign the respective blocks to
  each other in a top-down manner. 
  Recursive multisection as performed
  by
  ~\cite{Schulz2017a} models the hierarchy
  inherent in $G_p$ as a tree. The tree's fan-out then determines into
  how many blocks $G_c$ 
  needs to be partitioned.
Again, \mswap is complementary in that we do not need an actual
hierarchy of the topology.

\paragraph{Overview of our method.}
%
%
To the best of our knowledge, \mswap is the first
method to exploit the fact that many parallel topologies are partial cubes.
A partial cube is an isometric subgraph of a hypercube, see Section~\ref{sec:pc}
for a formal definition. This graph class includes all rectangular and cubic meshes, any such torus with
even extensions in each dimension, all hypercubes, and all trees.

$G_p$ being a partial cube allows us to label its vertices with
\emph{bitvectors} such that the distance between two vertices in $G_p$
amounts to the Hamming distance (number of different bits) between the
vertices' labels (see Sections~\ref{sec:pc} and~\ref{sec:labels_Gp}).
This will allow us to compute the contribution of an edge $e_a \in E_a$ to $\cc(\cdot)$ quickly.
The labels are then extended to labels for vertices in $G_a$
(Section~\ref{sec:labels_Ga}). More specifically, a label of a vertex
$v_a \in V_a$ consists of a left and a right part, where the left part
encodes the mapping $\mu(\cdot): V_a \mapsto V_p$ and the right part
makes the labels unique.
Note that it can be instructive to view the vertex labels of $V_a$, $l_a(\cdot)$, as a
recursive bipartitioning of $V_a$: the $i$-th leftmost bit of $l_a(v_a)$ defines
whether $v_a$ belongs to one or the other side of the corresponding cut in the $i$-th recursion level.
Vertices $u_a, v_a \in V_a$ connected by an edge with high weight should then be assigned labels that
share many of the bits in the left part.

%
\label{subsec:contribPlan}
%
 We extend the objective function $\cc(\cdot)$ by a factor that accounts for the labels' right part
 and thus for the uniqueness of the label, 
We do so without limiting the original objective that improves the labels'
left part (and thus  $\mu(\cdot)$) (Section~\ref{sec:extendObjFunc}).
Finally, we use the labeling of $V_a$ to devise a multi-hierarchical search
  method in which label swaps improve $\mu(\cdot)$. 
One can view this as a solution method for finding a $\cc(\cdot)$-optimal numbering of the vertices in $V_a$,
where the numbering is determined by the labels.

Compared to simple hill-climbing methods, we increase the local search space by employing very diverse hierarchies. These hierarchies result from random permutations of the label entries; they are imposed on the vertices of $G_p$ and built by label contractions performed digit by digit. This approach
  provides a fine-to-coarse view during the optimization process (Section~\ref{sec:multiSwap})
  that is orthogonal to typical multilevel graph partitioning methods.

\paragraph{Experimental results.}
\label{par:intro-exp-results}
As the underlying application for our experiments,  we assume parallel complex network analysis
on systems with distributed memory. Thus, in Section~\ref{sec:experiments}, we apply \mswap to further improve
mappings of complex networks onto partial cubes with 256
and 512 nodes. The initial mappings are computed using \kahip~\cite{sandersschulz13}, \scotch~\cite{Pellegrini07scotch} and \libtopomap~\cite{hoefler-topomap}.
To evaluate the quality results of \mswap,
  we use $\cc(\cdot)$, for which we observe
  a relative improvement from 6\% up to 34\%, depending on the choice of the initial mapping.
Running times for \mswap
  are on average 42\% faster than the running times for partitioning the graphs with \kahip.


%
%
\section{Partial cubes and their hierarchies} 
\label{sec:pc}

The graph-theoretical concept of a partial cube is central for this paper. Its definition
is based on the hypercube.
\begin{definition}[Hypercube]
  A $dim_H$-dimensional hypercube $H$ is the graph
  $H(V_H, E_H)$ with $V_H := \{0, 1\}^{dim_H}$ and $E_H := \{\{u_H, w_H\} \mid u_H, v_H \in V_H
  \mbox{~and~} d_h(u_H, v_H) = 1\}$, where $d_h(u_H, v_H)$ denotes the
  Hamming distance (number of different bits) between $u_H$ and $v_H$.    
\end{definition}
 More generally, $d_H(u_H, v_H) = d_h(u_H, v_H)$, \iec the (unweighted) shortest path length equals the Hamming distance in a hypercube. 

Partial cubes are isometric subgraphs of hypercubes, \iec the distance between any two nodes in a partial cube is the same as their distance in the hypercube. 
Put differently:
\begin{definition}[Partial cube]
A graph $G_p$ with vertex set $V_p$ is a partial cube of
dimension $dim_{G_p}$ if (i) there exists a labeling $l_p: V_p
\mapsto \{0, 1\}^{dim_{G_p}}$ such that $d_{G_p}(u_p, v_p) = d_h(l_p(u_p),
l_p(v_p))$ for all $u_p, v_p \in V_p$ and (ii) $dim_{G_p}$ is as small
as possible.
\label{def:partialCube}
\end{definition}

\begin{figure}[tb]
\begin{center}
\includegraphics[width=0.38\textheight]{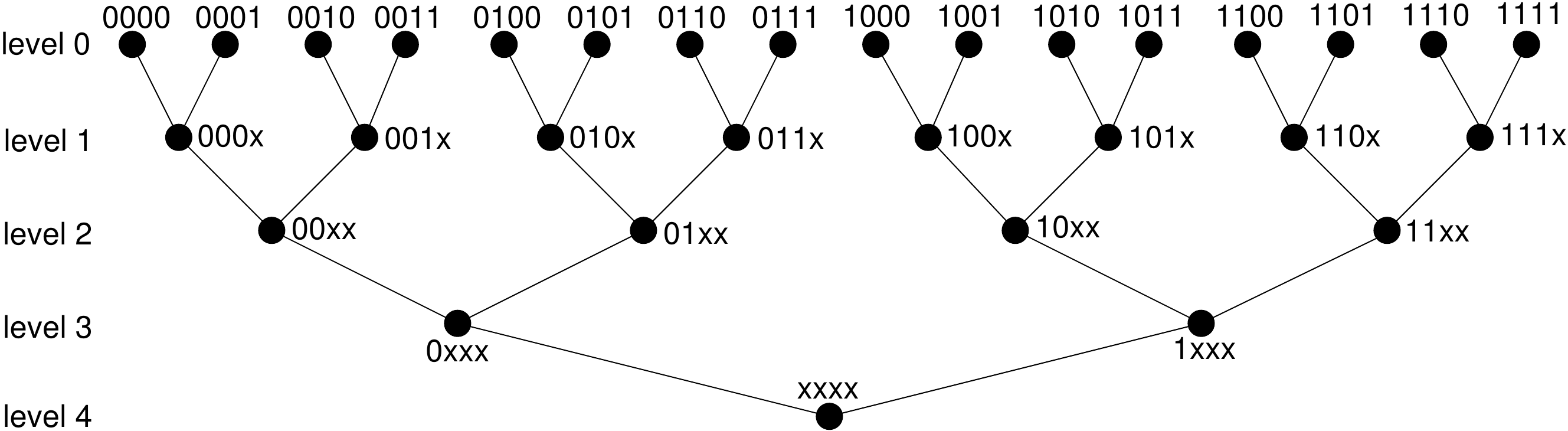}\\
~\\
\includegraphics[width=0.38\textheight]{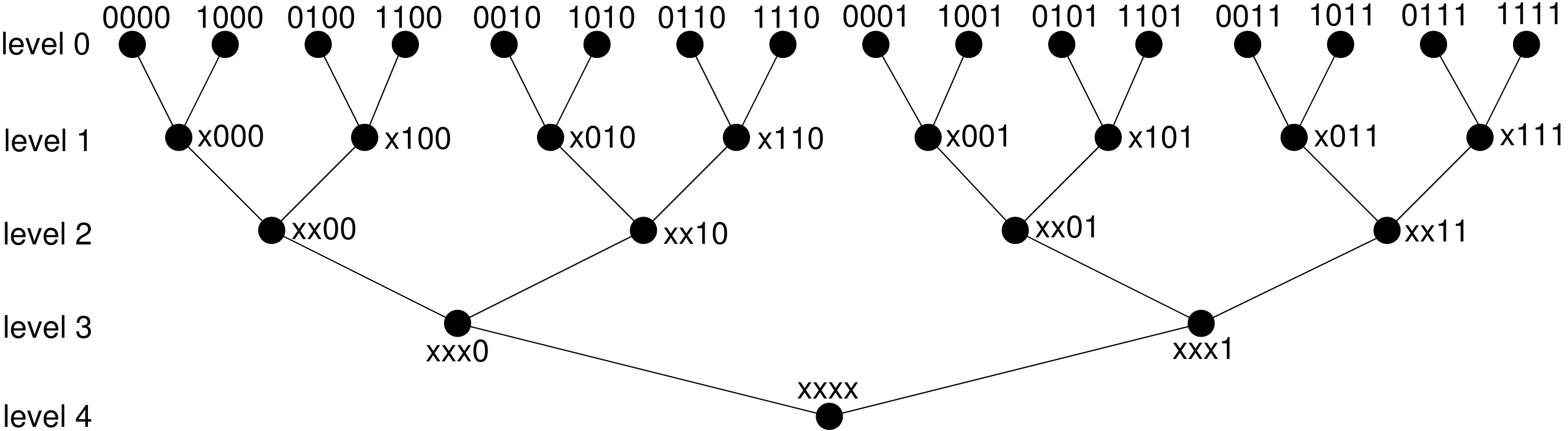}
\end{center}
\caption{\label{fig:twoHierarchies} {\small Two opposite hierarchies of the 4D
  hypercube. ``x'' means ``$0$'' or ``$1$''. Top: hierarchy $H_{\pi}$
  with $\pi = (1, 2, 3, 4)$. Bottom: hierarchy $H_{\pi}$ with $\pi =
  (4, 3, 2, 1)$.}}
\end{figure}

The labeling $l_p~:~V_p \mapsto \{0, 1\}^{dim_{G_p}}$ gives rise
  to hierarchies on $V_p$ as follows. For any permutation $\pi$ of
  $\{1, \dots, dim_{G_p}\}$, one can group the vertices in $V_p$ by
  means of the equivalence relations $\sim_{\pi, i}$, where $dim_{G_p}
  \geq i \geq 1$:
\begin{equation}
u_p \sim_{\pi, i} v_p :\Leftrightarrow \pi(l_p(u_p))[j] =
\pi(l_p(v_p))[j]
\end{equation}
for all $1 \leq j \leq i$ (where $l[i]$ refers to the $i$-th character in string $l$). 
As an example, $\sim_{id, i}$, where
  $id$ is the identity on $\{0, 1\}^{dim_{G_p}}$, gives rise to the
  partition in which $u_p$ and $v_p$ belong to the same part if and
  only if their labels agree at the first $i$ positions. More
  generally, for each permutation $\pi(\cdot)$ from the set
  $\Pi_{dim_{G_p}}$ of all permutations on $\{1, \dots, dim_{G_p}\}$,
  the equivalence relations $\sim_{\pi, i}$, $dim_{G_p} \geq i \geq
  1$, give rise to a hierarchy of increasingly coarse partitions
  $(\mathcal{P}_{dim_{G_p}}, \dots, \mathcal{P}_1)$. As an example,
  the hierarchies defined by the permutations $id$ and $\pi(j) :=
  dim_{G_p} + 1 - j$, $1 \leq j \leq dim_{G_p}$, respectively, are
  opposite hierarchies, see Figure~\ref{fig:twoHierarchies}.

\section{Vertex labels of processor graph}
\label{sec:labels_Gp}

In this section we provide a way to recognize if a graph $G_p$ is
a partial cube, and if so, we describe how a labeling on $V_p$
can be obtained in $\bigO(\vert E_p\vert^2)$ time.
To this end, we characterize partial cubes in terms of
(\emph{cut-sets} of) \emph{convex cuts}.

%
%
\begin{definition}[Convex cut] Let $G_p = (V_p, E_p)$
be a graph, and let $(V_p^0, V_p^1)$ be a cut of $G_p$. The cut
is called \emph{convex} if no shortest path from a vertex in $V_p^0$
[$V_p^1$] to another vertex in $V_p^0$ [$V_p^1$] contains a vertex of
$V_p^1$ [$V_p^0$].
\label{def:convexCut}
\end{definition}

\begin{figure}
\subfloat[Processor graph $G_p$]{\includegraphics[width=2.6cm]{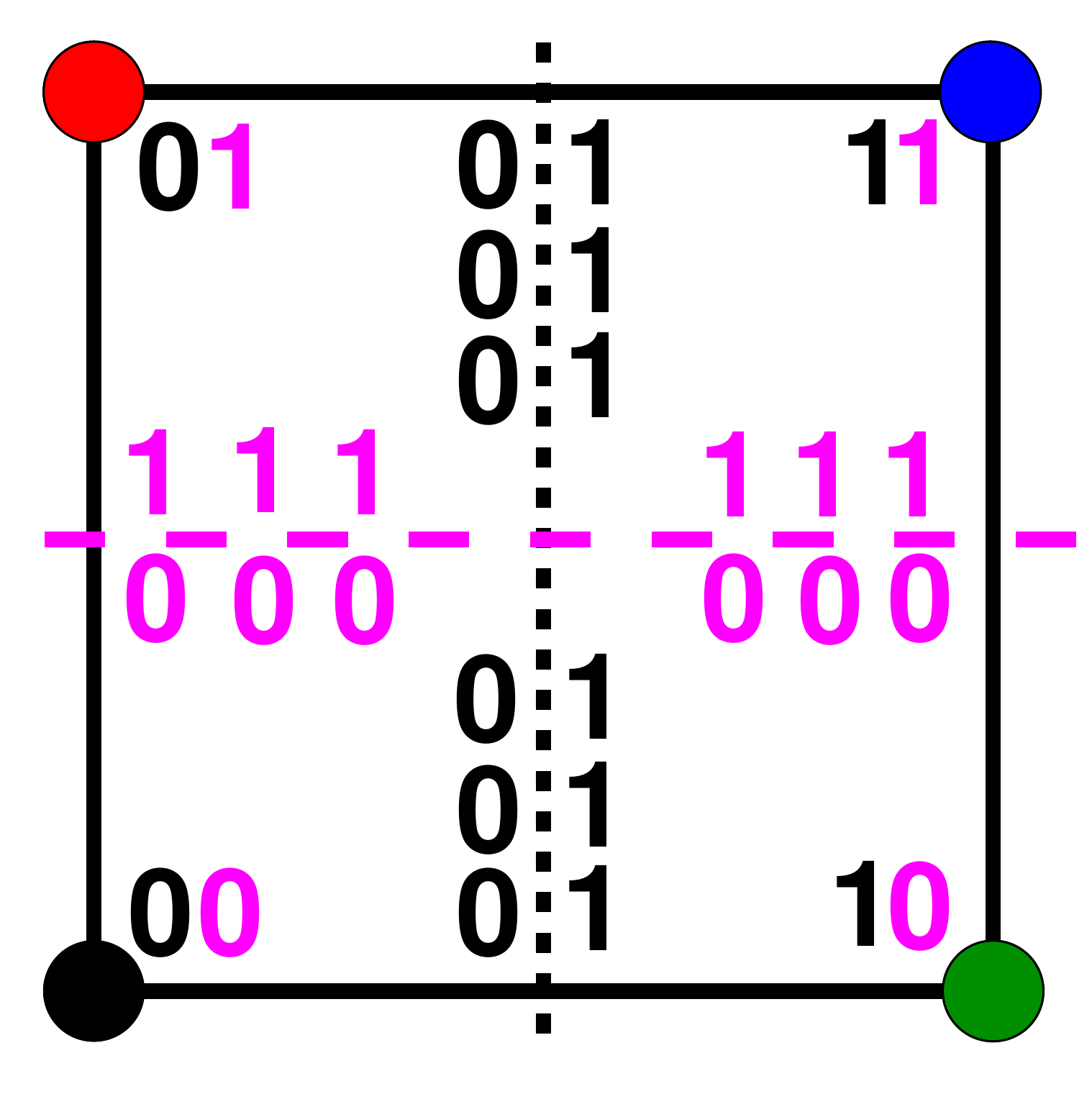}\label{a}}
\subfloat[Distances in $G_p$]{\includegraphics[width=2.6cm]{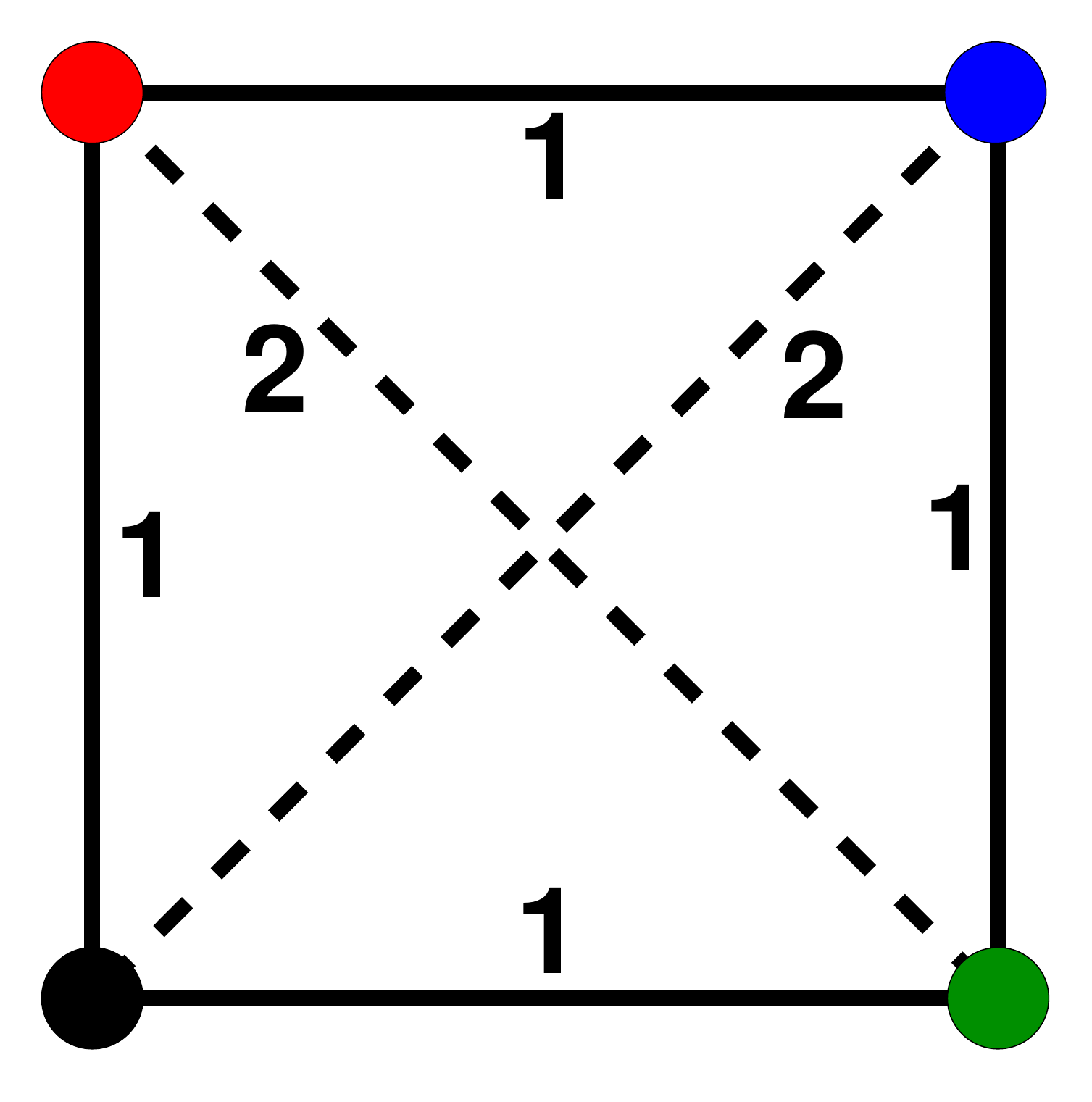}\label{b}}
\subfloat[Application graph $G_a$]{\includegraphics[width=2.6cm]{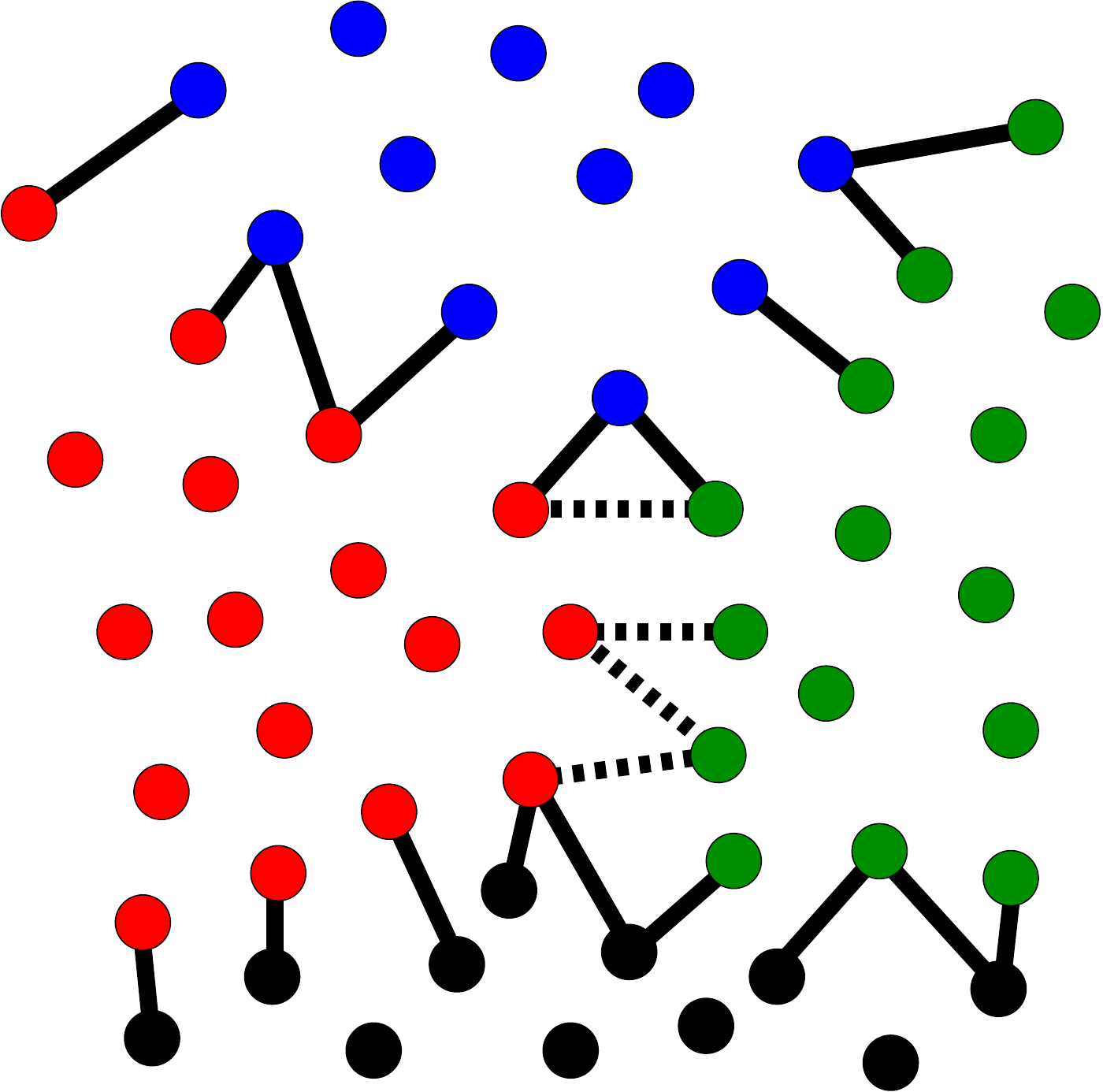} \label{c}}                                    
\caption{\label{fig:pc} {\small $G_p$ is a partial
    cube with two convex cuts: the dotted vertical cut and the
    dashed horizontal cut (\ref{a}). First [second] digit of vertex labels
    $l_p(\cdot)$ indicates position \wrt vertical [horizontal]
    cut. Distance between $u_p$ and $v_p$ in $G_p$ equals
    Hamming distance between $l_p(u_p)$ and $l_p(v_p)$ (\ref{b}). In \ref{c},
    a mapping $\mu(\cdot)$ from $V_a$ to $V_p$' is indicated by the
    colors. Communication across solid [dashed] edges requires 1 hop
    [2 hops] in $G_p$.}}
\end{figure}

The \emph{cut-set} of a cut $(V_p^0, V_p^1)$ of $G_p$ consists
  of all edges in $G_p$ with one end vertex in $V_p^0$ and the other
  one in $V_p^1$. Given the above definitions, $G_p$ is a partial cube if and only
  if (i) $G_p$ is bipartite and (ii) the cut-sets of $G_p$'s convex cuts partition
  $E_p$~\cite{Ovchinnikov2008a}. In this case, the equivalence relation behind the
    partition is the \djoko $\theta$~\cite{Ovchinnikov2008a}.
  Let $e_p = \{x_p, y_p\} \in E_p$. An edge $f_p$ is \djokoRelated to $e_p$
    if one of $f_p$'s end vertices is closer to $x_p$ than to $y_p$,
    while the other end vertex of $f_p$ is closer to $y_p$ than to
    $x_p$. Formally,
\begin{align*}
e_p~\theta~f_p&:\Leftrightarrow \vert f_p \cap W_{x_p,y_p} \vert =
\vert f_p \cap W_{y_p,x_p} \vert = 1 \mbox{~, where}\\
W_{x_p, y_p}&=\{w_p \in V_p \mid d_{G_p}(w_p,x_p) < d_{G_p}(w_p,y_p)\}.
\end{align*}
Consequently, no pair of edges on any shortest path of $G_p$ is
  \djokoRelated. 
In the following, we use the above definitions to find out whether $G_p$ is a partial cube
  and, if so, to compute a labeling $l_p~:~V_p \mapsto \{0,
  1\}^{dim_{G_p}}$ for $V_p$ according to Definition~\ref{def:partialCube}:
\begin{enumerate}
\item Test whether $G_p$ is bipartite (in asymptotic time $\bigO(\vert
  E_p \vert)$). If $G_p$ is not bipartite, it is not a partial cube. 
\item Pick an arbitrary edge $e_p^1$ and compute the edge set $E_p(e_p^1,
  \theta) := \{f_p \in E_p~\mid~f_p~\theta~e_p^1\}$.
\item Keep picking edges $e_p^2, e_p^3, \dots$ that are not contained
  in an edge set computed so far and compute the edge sets $E_p(e_p^2,
  \theta),$ $E_p(e_p^3, \theta)$, $\dots$. If there is an overlap with a
  previous edge set, $G_p$ is not a partial cube.
\item While calculating $E_p(e_p^j, \theta)$
  where $1 \leq j \leq dim_{G_p}$, set
\begin{equation}
  l_p[j](u_p) := \begin{cases}
    0, & \text{if $u_p \in W_{x_p^j,y_p^j}$}.\\
    1, & \text{otherwise}.
  \end{cases}
\end{equation}
For an example of such vertex labels see Figure~\ref{fig:pc}a.
\end{enumerate}

Assuming that all distances between node pairs are computed beforehand,
calculating $E_p(e_p^j,\theta)$ for $1 \leq j \leq dim_{G_p}$ and setting the
labeling $l_p(\cdot)$ (in steps $2$,$3$ and $4$)
take $\bigO(\vert E_p \vert)$ time. 
Detecting overlaps with already calculated edge sets takes $\bigO(\vert E_p\vert^2)$
time (in step $3$), which summarizes the time complexity of the proposed method.

Note that the problem has to be solved only
once for a parallel computer. Since $\vert E_p \vert = \bigO(\vert V_p \vert)$
for 2D/3D grids and tori -- the processor
graphs of highest interest in this paper -- and $\vert E_p \vert = \bigO(\vert V_p \vert \log{\vert V_p \vert})$
for all partial cubes,
our simple method is (almost) as fast, asymptotically, as the
fastest and rather involved methods that solve the problem.
Indeed, the fastest method to solve the problem takes time
$\bigO(\vert V_p \vert \vert E_p \vert)$~\cite{Imrich93a}. Assuming
that integers of at least $\log_2(\vert V_p \vert)$ bits can be stored
in a single machine word and that addition, bitwise Boolean
operations, comparisons and table look-ups can be performed on these
words in constant time, the asymptotic time is reduced to
$\bigO(\vert V_p \vert^2)$~\cite{Eppstein2011a}.

%
%
\section{Vertex labels of application graph}
\label{sec:labels_Ga}
Given an application graph $G_a=(V_a, E_a, \omega_a(\cdot))$, a
processor graph $G_p = (V_p, E_p)$ that is a partial cube and a
mapping $\mu: V_a \mapsto V_p$, the aim of this section is to define a
labeling $l_a(\cdot)$ of $V_a$. This labeling is then used later to
improve $\mu(\cdot)$ \wrt Eq.~(\ref{eq:objFuncCommCost2}) by swapping labels between
  vertices of $G_a$. In particular, for any $u_a, v_a \in V_a$, the
  effect of their label swap on $\mu(\cdot)$
  should be a function of their labels and those of their
  neighbors in $G_a$.
It turns out that being able to access the
  vertices of $G_a$ by their (unique) labels is crucial for the
  efficiency of our method. The following requirements on $l_a(\cdot)$
  meet our needs.

\begin{enumerate}
\item $l_a(\cdot)$ encodes $\mu(\cdot)$. 
\item For any two vertices $u_a, v_a \in G_a$, we can derive the
  distance between $\mu(u_a)$ and $\mu(v_a)$ from $l_a(u_a)$ and
  $l_a(v_a)$. Thus, for any edge $e_a$ of $G_a$, we can find out how
  many hops it takes in $G_p$ for its end vertices to
  exchange information, see Figure~\ref{c}.
\item The labels are unique on $V_a$.
\end{enumerate}
To compute such $l_a(\cdot)$, we first transport
  labeling $l_p(\cdot)$ 
  from $V_p$ to $V_a$ through $l_p(v_a) := l_p(\mu(v_a))$ for all $v_a \in
  V_a$. This new labeling $l_p:~V_a \mapsto \{0, 1\}^{dim_{G_p}}$
  already fulfills items 1) and 2). Indeed, item 1) holds, since
  labels are unique on $V_p$; item 2) holds,
  because $G_p$ is a partial cube. 
To fulfill item 3), we extend the labeling $l_p: V_a \mapsto
\{0,1\}^{dim_{G_p}}$ to a labeling $l_a: V_a \mapsto
\{0,1\}^{dim_{G_a}}$, where yet undefined $dim_{G_a}$ should exceed
$dim_{G_p}$ only by the smallest amount necessary to ensure that
$l_a(u_a) \neq l_a(v_a)$ whenever $u_a \neq v_a$. The gap $dim_{G_a} -
dim_{G_p}$ depends on the size of the largest part in the partition
induced by $\mu(\cdot)$.
\vspace{0.2cm}
\begin{definition}[$dim_{G_a}$]
\label{def:dimGa}
Let $\mu: V_a \mapsto V_p$ be a mapping. We set:
\begin{equation}
dim_{G_a} = dim_{G_a}(\mu) = dim_{G_p} + \max_{v_p \in V_p} \lceil \log_2 \vert \mu^{-1}(v_p) \vert \rceil.
\end{equation}
\end{definition}
%
For any $v_a \in V_a$, $l_a(v_a)$ is a bitvector of length $dim_{G_a}$;
its first $dim_{G_p}$ entries coincide with $l_p(v_a)$, see
above, and its last $dim_{G_a} - dim_{G_p}$ entries serve to
make the labeling unique. We denote the bitvector formed by the last
$dim_{G_a} - dim_{G_p}$ entries by $l_e(v_a)$. Here, the subscript $e$
in $l_e(\cdot)$ stands for ``extension''. To summarize,
\begin{equation}
l_a(v_a) = l_p(v_a) \circ l_e(v_a) \mbox{~~for all $v_a \in V_a$}, \text{where}
\end{equation}
$\circ$ stands for concatenation. Except for temporary permutations of
the labels' entries, the set $\mathcal{L} := l(V_a)$ of labels
will remain the same. A label swap between $u_a$ and $v_a$
  alters $\mu(\cdot)$ if and only if $l_p(u_a) \neq l_p(v_a)$. The
  balance of the partition of $V_a$, as induced by $\mu(\cdot)$, is
  preserved by swapping the labels of $V_a$.

Computing $l_e(\cdot)$ is straightforward. First, the vertices in
each $\mu^{-1}(v_p)$, $v_p \in V_p$, are numbered from $0$ to $\vert
\mu^{-1}(v_p) \vert - 1$. Second, these decimal numbers are then interpreted as
bitvectors/binary numbers. Finally, the entries of the corresponding bitvectors are
shuffled, so as to provide a good random starting point for the improvement
of $\mu$, see Lemma~\ref{lemma:benefit} in
Section~\ref{sec:extendObjFunc}.

%
%
\section{Extension of the objective function}
\label{sec:extendObjFunc}
%

  Given the labeling of vertices in $V_a$, \iec $l_a(\cdot) = l_p(\cdot) \circ l_e(\cdot)$,
  it is easy to see that solely $l_p(\cdot)$ determines the value
  of $\cc(\cdot)$. 
(This fact results from $l_p(\cdot)$ encoding the distances between vertices in $G_p$.) 
On the other hand, due to the uniqueness of the labels $l_a(\cdot)$, $l_e(\cdot)$ restricts $l_p(\cdot)$:
  $l_e(u_a) = l_e(v_a)$ implies $l_p(u_a) \neq l_p(v_a)$, \iec that $u_a$ and $v_a$
  are mapped to different PEs.
  
  

The plan of the following is to ease this restriction by
  avoiding as many cases of $l_e(u_a) = l_e(v_a)$ as possible. To this
  end, we incorporate $l_e(\cdot)$ into the objective function, thus replacing
  $\cc(\cdot)$ by a modified one.
Observe that $l_p(\cdot)$ and $l_e(\cdot)$ give
rise to two disjoint subsets of $E_a$, \ie subsets of edges, the two end vertices
of which agree on $l_p(\cdot)$ and $l_e(\cdot)$, respectively:
\begin{align*}
E_a^p = E_a^p(l_a) &:= \{e_a = \{u_a, v_a\} \in E_a \mid l_p(u_a) = l_p(v_a)\},\\
E_a^e = E_a^e(l_a) &:= \{e_a = \{u_a, v_a\} \in E_a \mid l_e(u_a) = l_e(v_a)\}.
\end{align*}
In general these two sets do not form a partition of $E_a$, since there can be edges whose end vertices disagree both on $l_p(\cdot)$ and on $l_e(\cdot)$.
With $h(\cdot, \cdot)$ denoting the Hamming distance, optimizing
$\cc(\cdot)$ in Eq.~(\ref{eq:objFuncCommCost2}) can be rewritten as
follows. Find
\begin{align}
\label{eq:objFuncCommCost3}
l_a^* &:= \argmin\limits_{\substack{l_a: V_a \mapsto
    \mathcal{L}\\\mbox{\tiny $l_a$~bijective}}} \cc(l_a)
\mbox{, where}\\
\label{eq:objFuncCommCost4}
\cc(l_a) &:= \sum_{\substack{e_a \in E_a \setminus E_a^p(l_a)\\{\mbox{\tiny $e_a=\{u_a,v_a\}$}}}}
\hspace{-0.3cm} \omega_a(e_a)~h(l_p(u_a),l_p(v_a)).
\end{align}

For all $e_a = \{u_a, v_a\} \in E_a^e$ it follows that $l_p(u_a) \neq l_p(v_a)$, implying that $u_a$ and $v_a$ are mapped to different PEs. Thus, any
 edge $e_a \in E_a^e(l_a)$ increases the value of $\cc(\cdot)$ with a damage of
%
%
\begin{equation}
\label{eq:damage}
\omega_a(u_a,v_a)~h(l_p(u_a),l_p(v_a)) > 0.
\end{equation}

This suggests that reducing $E_a^e$ may be good for minimizing
  $\cc(\cdot)$. The crucial question here is whether reducing $E_a^e$
  can obstruct our primary goal, \iec growing $E_a^p$ (see
  Eq.~(\ref{eq:objFuncCommCost4})). Lemma~\ref{lemma:benefit} below
  shows that this is not the case, at least for perfectly balanced
  $\mu(\cdot)$. A less technical version of Lemma~\ref{lemma:benefit}
  is the following: If $\mu(\cdot)$ is perfectly balanced, then two
  mappings, where one has bigger $E_a^p$ and one has smaller
  $E_a^e$ (both \wrt set inclusion), can be combined to a third mapping
  that has the bigger $E_a^p$ and the smaller $E_a^e$. The third
  mapping provides better prospects for minimizing $\cc(\cdot)$ than
  the mapping from which it inherited big $E_a^p$, as, due to
  smaller $E_a^e$, the third mapping is less constrained by the
  uniqueness requirement:
\vspace{0.2cm}
\begin{lemma}[Reducing $E_a^e$ compatible with growing $E_p^e$]
\label{lemma:benefit}
Let $\mu: V_a \mapsto V_p$ be such that $\vert \mu^{-1}(u_p)
  \vert = \vert \mu^{-1}(v_p) \vert$ for all $u_p, v_p \in V_p$
  (perfect balance). Furthermore, let $l_a(\cdot)$ and $l_a'(\cdot)$
  be bijective labelings $V_a \mapsto \mathcal{L}$ that correspond
  to $\mu(\cdot)$ as specified in Section~\ref{sec:labels_Ga}. Then,
  $E_a^p(l_a) \supseteq E_a^p(l_a')$ and $E_a^e(l_a') \subseteq
  E_a^e(l_a)$ implies that there exists $l_a^*$ that also
  corresponds to $\mu(\cdot)$ with (i) $E_a^p(l_a^*) = E_a^p(l_a)$
  and (ii) $E_a^e(l_a^*) = E_a^e(l_a')$.
\end{lemma}
\vspace{0.2cm}
\begin{proof}
Set $l_a^*(\cdot) := l_a^p(\cdot) \circ l_a^e(\cdot)$. Then,
  $l_a^*(\cdot)$ fulfills (i) and (ii) in the claim, and the first
  $dim_{G_p}$ entries of $\mu(\cdot)$ specify $\mu(\cdot)$. If remains
  to show that the labeling $l_a^*(\cdot)$ is unique on $V_a$. This
  is a consequence of (a) $l_a(\cdot)$ being unique and (b)
  $E_a^e(l_a^*) = E_a^e(l_a') \subseteq E_a^e(l_a)$.
\end{proof}
\vspace{0.2cm}
%

In practice we usually do not have perfect balance. Yet, the balance
is typically low, \eg $\epsilon = 0.03$. Thus, we still
expect that having small $E_a^e(l_a)$ is beneficial for minimizing
$\cc(\cdot)$.

Minimization of the damage to $\cc(\cdot)$ from edges in $E_a^e(l_a)$,
see Eq.~(\ref{eq:damage}), amounts to maximizing the \emph{diversity}
of the label extensions in $G_a$. Formally, in order to diversify, we want to find
\begin{align}
\label{eq:divers1}
l_a^* &:= \argmax\limits_{\substack{l_a: V_a \mapsto
    \mathcal{L}\\\mbox{\tiny $l_a$~bijective}}} \divers(l_a)
\mbox{, where}\\
\label{eq:divers2}
\divers(l_a) &:= \sum_{\substack{e_a \in E_a \setminus E_a^e(l_a)\\{\mbox{\tiny $e_a=\{u_a,v_a\}$}}}}
\hspace{-0.3cm} \omega_a(e_a)~h(l_e(u_a),l_e(v_a)).
\end{align}

We combine our two objectives, \iec minimization of $\cc(\cdot)$ and
maximization of $\divers(\cdot)$, with the objective
function $\ccd(\cdot)$:
\begin{align}
\label{eq:objFuncCommCost5}
l_a^* &:= \argmin\limits_{\substack{l_a: V_a \mapsto
    \mathcal{L}\\\mbox{\tiny $l_a$~bijective}}} \ccd(l_a)
\mbox{, where}\\
\label{eq:objFuncCommCost6}
\ccd(l_a) &:= \cc(l_a) - \divers(l_a).
\end{align}

%
%
\section{Multi-hierarchical label swapping}
\label{sec:multiSwap}
After formulating the mapping problem as finding an optimal labeling for the vertices in $V_a$, we can now turn our attention to \emph{how} to find such a labeling -- or at least a very good one.
Our algorithm is meant to \emph{improve} mappings and resembles a key ingredient of the
classical graph partitioning algorithm by Kernighan and Lin
(KL)~\cite{Kernighan1970a}: vertices of $G_a$ swap their membership to
certain subsets of vertices. Our strategy differs from KL in that we
rely on a rather simple local search which is, however, performed on
multiple (and very diverse) random hierarchies on $G_a$. These
  hierarchies are oblivious to $G_a$'s edges and correspond to
  recursive bipartitions of $G_a$, which, in turn, are extensions of
  natural recursive bipartitions of $G_p$.

\subsection{Algorithm \mswap}
Our algorithm, see procedure \mswap in Algorithm~\ref{algo:mswap},
takes as input (i) an application graph $G_a$, (ii) a processor graph $G_p$ with the
partial cube property, (iii) an initial mapping $\mu: V_a \mapsto V_p$
and (iv) the number of hierarchies, $N_H$. $N_H$
controls the tradeoff between running time and the quality of the results.  The output
of \mswap consists of a bijective labeling $l_a: V_a \mapsto\mathcal{L}$ such that $\ccd(l_a)$ is low
(but not necessarily optimal).
Recall from Section~\ref{sec:intro} that requiring $\mu(\cdot)$ as input is no major limitation.
An initial bijection $l_a(\cdot)$ representing this $\mu(\cdot)$ is found in
lines 1, 2 of Algorithm~\ref{algo:mswap}.

In lines 3 through 21 we take $N_H$ attempts to improve $l_a(\cdot)$, where each
attempt uses another hierarchy on $\{0, 1\}^{dim_{G_a}}$.
Before the new hierarchy is built, the old labeling is
saved in case the label swapping in the new hierarchy turns out
to be a setback \wrt $\ccd(\cdot)$ (line 4). This may occur, since
the gain \wrt $\ccd(\cdot)$ on a coarser level of a hierarchy is only
an estimate of the gain on the finest level (see
below). The hierarchy and the current mapping are encoded by (i) a
sequence of graphs $G_a^i = (V_a^i, E_a^i, \omega_a^i)$, $1 \leq
  i \leq dim_{G_a}$, with $G_a^1 = G_a$, (ii) a sequence of
labelings $l_a^i : V_i \mapsto \{0, 1\}^{dim_{G_a^i}}$ and
(iii) a vector, called $\parent$, that provides the hierarchical
relation between the vertices.
From now on, we interpret vertex labels as integers whenever
convenient. More precisely, an integer arises from a label, \iec a
bitvector, by interpreting the latter as a binary number.

\begin{figure}[tb]
\centering
\includegraphics[height=0.18\textheight]{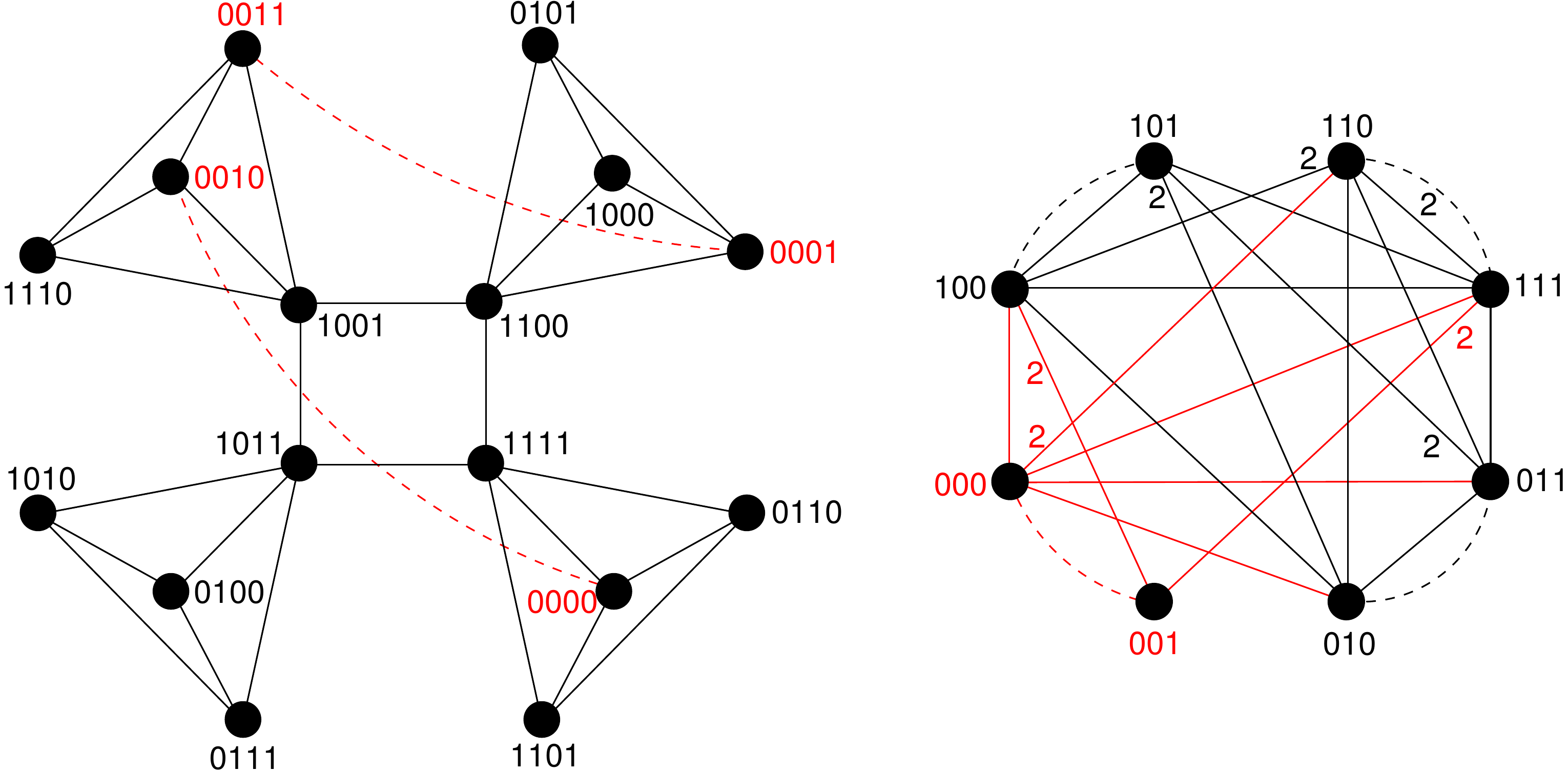}
\caption{\label{fig:gain1} {\small Graphs $G_a^1$ on level 1 of the
    hierarchy and $G_a^2$ on level 2 ($G_a^2$ arises from $G_a^1$
    through contractions controlled by the labels) are shown on the
    left and right, respectively. The first [last] two digits of the
    labels on $G_a^1$'s vertices indicate $l_p(\cdot)$ [$l_e(\cdot)$], respectively.
    Swapping labels $000$ and $001$ on
    $G_a^2$ yields a gain of $1$ in diversity (see
    Eq.~(\ref{eq:divers2})). The corresponding swaps in $G_a^1$ are
    indicated by the dashed red lines on the left.}}
\end{figure}
  
In lines 6 and 7, the entries of the vertex labels of $G_a$ are permuted according to a
random permutation $\pi(\cdot)$. The construction of
the hierarchy (lines 9 through 14) goes hand in hand with label
swapping (lines 10-12) and label coarsening (in line 13). The
function $\contract(\cdot, \cdot, \cdot)$ contracts any pair of
vertices of $G_a^{i-1}$ whose labels agree on all but the last digit,
thus generating $G_a^i$. The same function
also cuts off the last digit of $l_a^{i-1}(v)$ for all $v \in V_a^{i-1}$,
and creates the labeling $l_a^i(\cdot)$ for the next
coarser level. Finally, $\contract(\cdot, \cdot, \cdot)$ builds the
vector $\parent$ (for encoding the hierarchy of the vertices).
In line 15, the call to $\assemble()$ derives a new labeling
$l_a(\cdot)$ from the labelings $l_a^i(\cdot)$ on the levels $1 \leq i \leq dim_{G_a} - 1$ of a hierarchy
(read more in Section~\ref{sub:assemble}).
The permutation of $l_a$'s entries is undone in line
16, and $l_a(\cdot)$ is kept only if better than $l_{old}(\cdot)$, see
lines 17 to 19. Figure~\ref{fig:gain1} depicts a snapshot of
\mswap on a small instance.

\begin{figure*}[ttt!]
  \begin{minipage}[t]{5.15in}
    {\small
  \begin{algorithm}[H]
    \caption{\small Procedure \mswap$(G_a, G_p, \mu(\cdot), N_H)$ returns
      a bijection $l_a: V_a \mapsto \{0, 1\}^{dim_{G_a}}$ with a low value
      of $\cc(l_a)$.}
    \label{algo:mswap}
    \begin{algorithmic}[1]
      \State Find a labeling $l_p(\cdot)$ of $V_p$'s vertices, as described in Section~\ref{sec:pc}
      \State Using $\mu(\cdot)$, extend $l_p(\cdot)$ to a labeling $l_a(\cdot)$ of
      $V_a$'s vertices, as described in Section~\ref{sec:labels_Ga}
      \For{$N' = 1 \dots, N_H$}
      \State $l_{old}(\cdot) \gets l_a(\cdot)$
      \Comment just in case $l_a(\cdot)$ gets worse \wrt $\ccd(\cdot)$
      \State $\parent \gets []$
      \Comment $\parent$ will encode the hierarchy of the vertices
      \State Pick a random permutation $\pi: \{1, \dots, dim_{G_a}\} \mapsto \{1, \dots, dim_{G_a}\}$
      \State $l_a(\cdot) \gets \pi(l_a(\cdot))$
      \State $G_a^1 \gets G_a$, $l_a^1(\cdot) \gets l_a(\cdot)$
      \For{$i = 2, \dots, dim_{G_a} - 1$}
      \ForAll{$u, v \in G_a^{i-1}$ with $l_a^{i-1}(u) / 2 =
        l_a^{i-1}(v) / 2$}
      \Comment only least sig. digit differs
      \State Swap labels $l_a^{i-1}(u)$ and $l_a^{i-1}(v)$ if this decreases
      $\ccd(l_a^{i-1})$ on $G_a^{i-1}$. 
      \EndFor
      \State $(G_a^i, l_a^i, \parent) \gets \contract(G_a^{i - 1}, l_a^{i-1}, \parent)$
      \EndFor
      \State $l_a(\cdot) \gets \assemble(G_a^1, \dots, G_a^{dim_{G} - 1}, l_a^1,
      \dots, l_a^{dim_{G_a}-1}, \parent)$
      \State $l_a(\cdot) \gets \pi^{-1}(l_a(\cdot))$
      \If{$\ccd(l_a) > \ccd(l_{old})$}
      \State $l_a(\cdot) \gets l_{old}(\cdot)$
      \EndIf
      \EndFor
      \State \Return $l_a(\cdot)$
    \end{algorithmic}
  \end{algorithm}
  }
\end{minipage}
\hfill
\end{figure*}

\begin{figure*}[ttt!]
  \begin{minipage}[t]{5.15in}
    {\small
    \begin{algorithm}[H]
\caption{\small Function $\assemble(G_a^1, \dots, G_a^{dim_{G} - 1},
  l_a^1, \dots, l_a^{dim_{G_a}-1}, \parent)$ returns a new labeling
  $l_a^1(\cdot)$ of the vertices of $G_a^1 = G_a$ and thus a new labeling
  $l_a(\cdot)$ of $G_a$'s vertices.}
\label{algo:assemble}
\begin{algorithmic}[1]
    \ForAll{$v_1 \in V_a^1$}
        \State $l_a^1(v_1) \gets l_a^1(v_1) \modulo 2$
        \Comment{Write least significant digit}
        \State $\oldParent \gets v_1$
        \State $i \gets 1$
        \While{$i < dim_{G_a^1}$}
            \Comment Write digits $2, \dots, \dim_{G_1} - 1$
            \State $\newParent \gets \parent(\oldParent)$
            \State $i \gets i + 1$
            \State $\newParentLabel \gets l_a^i(\newParent)$
            \State $\prefLabel \gets l_a^1(v_1) + (\newParentLabel \ll
            (i-1))$ \Comment Preferred $i$ least sig. digits
            \If{$\exists~w \in V_1$ with $l_a^1(w) \modulo 2^i =
              \prefLabel$} \Comment Part of existing label?
                \State $l_a^1(v_1) \gets l_a^1(v_1) + ((\newParentLabel \modulo 2) \ll (i-1))$
                \Comment Write preferred digit
            \Else
                \State $l_a^1(v_1) \gets l_a^1(v_1) + ((1 - (\newParentLabel \modulo 2)) \ll (i-1))$
                \Comment Write other digit
            \EndIf
            \State $\oldParent \gets \newParent$
        \EndWhile
        \If{$l_a^1(v_1) \geq 1 \ll (dim_{G_1} - 1)$}
            \State $l_a^1(v_1) \gets l_a^1(v_1) + (1 \ll (dim_{G_1} -
            1))$
            \Comment{Write most significant digit}
        \EndIf
    \EndFor
\State \Return $l_a^1(\cdot)$
\end{algorithmic}
    \end{algorithm}
    }
\end{minipage}
\hfill
\end{figure*}
  
\subsection{Function $\assemble()$}
\label{sub:assemble}
Function $\assemble()$ (Algorithm~\ref{algo:assemble}) in line 15 of
\mswap turns the hierarchy of graphs $G_a^i$ into a new labeling
$l_a(\cdot)$ for $G_a$, digit by digit, using labels $l_a^i(\cdot)$ (in Algorithm~\ref{algo:assemble}, ``$\ll i$''
denotes a left shift by $i$ digits). The least and the most significant digit
of any $l_a^1(v_1)$ are inherited from $l_a(\cdot)$ (line 7 of
  Algorithm~\ref{algo:mswap}) and do not change (lines 2, 17 and 18). The remaining digits 
are set in the loop from line 5 to 16. Whenever possible, digit
$i$ of $l_a^1(v_1)$ is set to the last digit of the parent of $v_1$ (= preferred digit)
on level $i$, see lines 9, 11.  This might, however, lead to a label that
is not in $l_a^1(V_a^1)$ any more, which would change the set of labels and may violate
the balance constraint coming from $\mu(\cdot)$.
To avoid such balance problems, we take the last digit of $l_a^i(v_a)$
if possible (in lines 9-11) or, if not, we switch to the (old) inverted digit, see
line 13. Since $G_a^1 = G_a$, new $l_a^1(\cdot)$ on $G_a^1$ can be
taken as new $l_a(\cdot)$ on $G_a$, see line 18 in
Algorithm~\ref{algo:mswap}.

\subsection{Running time analysis}
The expected asymptotic running time of function $\assemble()$ is
$\bigO(\vert V_a \vert \cdot dim_{V_a})$. Here, ``expected'' is due to
the condition in line 10 that is checked in \emph{expected}
constant time. (We use a hashing-based {\texttt C++ std::unordered\_map} to find
  a vertex with a certain label. A plain array would be too
  large for large grids and tori, especially if the blocks are large,
  too.) For Algorithm~\ref{algo:mswap}, the expected running time 
is dominated by the loop between lines 9 and 14. The loop between
lines 10 and 12 takes amortized expected time $\bigO(\vert E_a 
\vert)$ (``expected'', because we have to go from the labels to the
vertices and ``amortized'', because we have to check the neighborhoods
of all $u, v$). The contraction in line 13 takes time $\bigO(\vert E_a \vert)$, too.
Thus, the loop between lines 9 and 14 takes time $\bigO(\vert E_a \vert dim_{G_a})$.
In total, the expected running time of Algorithm~\ref{algo:mswap} is $\bigO(N_H~\vert E_a\vert~dim_{G_a})$.

An effective first step toward a parallel version of our
  algorithm would be simple loop parallelization in lines 10-12 of
  Algorithm~\ref{algo:mswap}. 
  To avoid stale data, label accesses need to be coordinated.


%
%
\section{Experiments}
\label{sec:experiments}

\subsection{Description of experiments}
\label{subsec:setup}
In this section we specify our test instances, our experimental setup
and the way we evaluate the computed mappings.

The application graphs are the 15 complex networks used
by Safro \etal~\cite{Safro2012a} for partitioning experiments and
in~\cite{Glantz2015c} for mapping experiments, see
Table~\ref{tab:complex}. Regarding the processor graphs, we follow
loosely current architectural trends.  Several leading supercomputers
have a torus architecture~\cite{7161509}, and grids (= meshes)
experience rising importance in emerging multicore
chips~\cite{5999870}. As processor graphs $G_p = (V_p, E_p)$ we
therefore use a 2DGrid($16 \times 16$), a 3DGrid($8 \times 8 \times
8$), a 2DTorus($16 \times 16$), a 3DTorus($8 \times 8 \times 8$) and,
for more theoretical reasons, an $8$-dimensional hypercube.
In our experiments, we set the number of hierarchies ($N_H$) for \mswap
to 50 and whenever is needed for partitioning/mapping with state-of-the-art tools,
the load imbalance is set to 3\%.
All computations are based on sequential C++ code.
Each experiment is executed on a node with two Intel XeonE5-2680 processors (Sandy Bridge) at 2.7 GHz equipped with 32 RAM
and 8 cores per processor.



{\footnotesize
  \begin{table*}[!b]
\caption{{\small Complex networks used for benchmarking.}}
\label{tab:complex}
\begin{center}
\begin{tabular}{ l | r | r | c }
Name & \#vertices & \#edges & Type\\ \hline \hline
p2p-Gnutella          & \numprint{6405}   & \numprint{29215}    & file-sharing network\\\hline
PGPgiantcompo         & \numprint{10680}  & \numprint{24316}    & largest connected component in network of PGP users\\\hline
email-EuAll           & \numprint{16805}  & \numprint{60260}    & network of connections via email\\\hline
as-22july06           & \numprint{22963}  & \numprint{48436}    & network of internet routers \\\hline
soc-Slashdot0902      & \numprint{28550}  & \numprint{379445}   & news network\\\hline
loc-brightkite\_edges & \numprint{56739}  & \numprint{212945}   & location-based friendship network\\\hline
loc-gowalla\_edges    & \numprint{196591} & \numprint{950327}   & location-based friendship network \\\hline
citationCiteseer      & \numprint{268495} & \numprint{1156647}  & citation network\\\hline
coAuthorsCiteseer     & \numprint{227320} & \numprint{814134}   & citation network\\\hline
wiki-Talk             & \numprint{232314} & \numprint{1458806}  & network of user interactions through edits\\\hline
coAuthorsDBLP         & \numprint{299067} & \numprint{977676}   & citation network\\\hline
web-Google            & \numprint{356648} & \numprint{2093324}  & hyperlink network of web pages\\\hline
coPapersCiteseer      & \numprint{434102} & \numprint{16036720} & citation network\\\hline
coPapersDBLP          & \numprint{540486} & \numprint{15245729} & citation network\\\hline
as-skitter            & \numprint{554930} & \numprint{5797663}  & network of internet service providers\\\hline
\end{tabular}
\end{center}
\end{table*}
}


\paragraph*{Baselines.}
For the evaluation, we use four different experimental cases ($\cone$ to $\cfour$), each of which
assumes a different initial mapping $\mu_1(\cdot)$
as an input to \mswap (Algorithm~\ref{algo:mswap}).
The different cases shall measure the improvement by \mswap
compared to different standalone mapping algorithms.
In the following, we describe how we obtain the initial mappings $\mu_1(\cdot)$ for each case separately.

In $\cone$ we compare the improvement of \mswap on an initial mapping produced by \scotch. For that, we use the generic mapping routine
of \scotch with default parameters. It returns a mapping $\mu_1(\cdot)$ of a given graph using a dual recursive bipartitioning algorithm.




In $\ctwo$ we use the $\initial$ mapping that maps
block $i$ of the application graph (or vertex $i$ of the communication
graph $G_c$) to node $i$ of the processor graph $G_p$, $1 \leq i \leq
\vert G_c \vert = \vert G_p \vert$.
$\initial$ receives its solution from the initial partition
computed with \kahip. This approach benefits from spatial locality in the partitions,
so that $\initial$ often yields surprisingly good
solutions~\cite{Glantz2015c}.

In $\cthree$ we use a mapping algorithm named $\greedyallcM$ that has been previously proposed by a subset of the authors (implemented on top of \kahip).
$\greedyallcM$ is an improvement of a previous
greedy algorithm~\cite{Brandfass2013372} and is the best performing algorithm in~\cite{Glantz2015c}. It builds on the
idea of increasing a mapping by successively adding assignments $v_c
\rightarrow v_p$ such that (a) $v_c \in G_c$ has maximal communication volume
with one or all of the already mapped vertices of $G_c$ and (b) $v_p \in G_p$
has minimal distance to one or all of the already mapped vertices of
$G_p$.

Finally, we compare against \libtopomap~\cite{hoefler-topomap}, a state-of-the-art mapping tool that includes multiple mapping algorithm.
More precisely we use the algorithm whose general idea follows
the construct method, in~\cite{Brandfass2013372} .
Subset of the authors has previous implemented the above algorithm on top of the \kahip tool (named $\greedymin$).
As a result, and in order to accommodate comparisons with $\ctwo$, $\cthree$
$\greedymin$ is used as the mapping algorithm for the experimental case $\cfour$.

\paragraph*{Labeling.}
  Once the initial mappings $\mu_1(\cdot)$ are calculated, we need to perform two more steps
  in order to get an initial labeling $l_a(\cdot)$:
\begin{enumerate}
\item We compute a labeling $l_p~:~V_p \mapsto \{0,
  1\}^{dim_{G_p}}$, where $l_p(\cdot)$ and $dim_{G_p}$ fulfill the
  conditions in Definition~\ref{def:partialCube}. In particular,
  $d_{G_p}(u_p, v_p) = d_h(l_p(u_p), l_p(v_p))$ for all $u_p, v_p \in
  V_p$, where $d_h(\cdot, \cdot)$ denotes the Hamming distance. Due to
  the sparsity of our processor graphs $G_p$ (grids, tori,
  hypercubes), we use the method outlined in
  Section~\ref{sec:labels_Gp}.
\item We extend the labels of $G_p$'s vertices to labels of
  $G_a$'s vertices as described in Section~\ref{sec:labels_Ga}.
 \end{enumerate}


  Then, for each experimental case, \mswap is given the initial mapping $\mu_1(\cdot)$
  and it generates a new mapping $\mu_2(\cdot)$.
  Here, we compare the quality of mapping $\mu_2(\cdot)$ to $\mu_1(\cdot)$
  in terms of our main objective function $\cc(\cdot)$,
  but we also provide results for the edgecut metric and for the
  running times.

\paragraph*{Metrics and parameters.}
Since \scotch, \kahip and \mswap have randomized components, we run each experiment
5 times. Over such a set of 5 repeated experiments we compute the
minimum, the arithmetic mean and the maximum of \mswap's running time
($T$), edge cut ($Cut$) and communication costs $\cc(\cdot)$ ($Co$).
Thus we arrive at the values
$T_{min}$, $T_{mean}, \dots, Co_{max}$ (9 values for each combination
of $G_a$, $G_p$, for each experimental case $\cone$ to $\cfour$). Each of
these values is then divided by the min, mean, and max value \emph{before}
the improvements by \mswap, except the running time of \mswap,  which
is divided by the partitioning time of \kahip for $\ctwo$,$\cthree$,$\cfour$ and by the mapping time of \scotch for $\cone$.
Thus, we arrive at 9 quotients $qT_{min}, \dots, qCo_{max}$. Except for the terms involving
running times, a quotient smaller than one means that \mswap was
successful.
Next we form the geometric means of the 9 quotients over the
application graphs of Table~\ref{tab:complex}. Thus we arrive at 9
values $qT^{gm}_{min}, \dots, qCo^{gm}_{min}$ for any combination of
$G_p$ and any experimental case.
Additionally, we calculate the geometric standard deviation as an indicator
of the variance over the normalized results of the application graphs.  
\subsection{Experimental Results}
\label{subsec:results}
The detailed experimental results regarding quality metrics are displayed in Figures~\ref{c1} through~\ref{c4} (one for each experimental case),
while the running time results are given in Table~\ref{tab:extra_social_initial_1}.
Here is our summary and interpretation:

\begin{figure*}
\subfloat[]{
\centering
\includegraphics[width=0.45\textwidth]{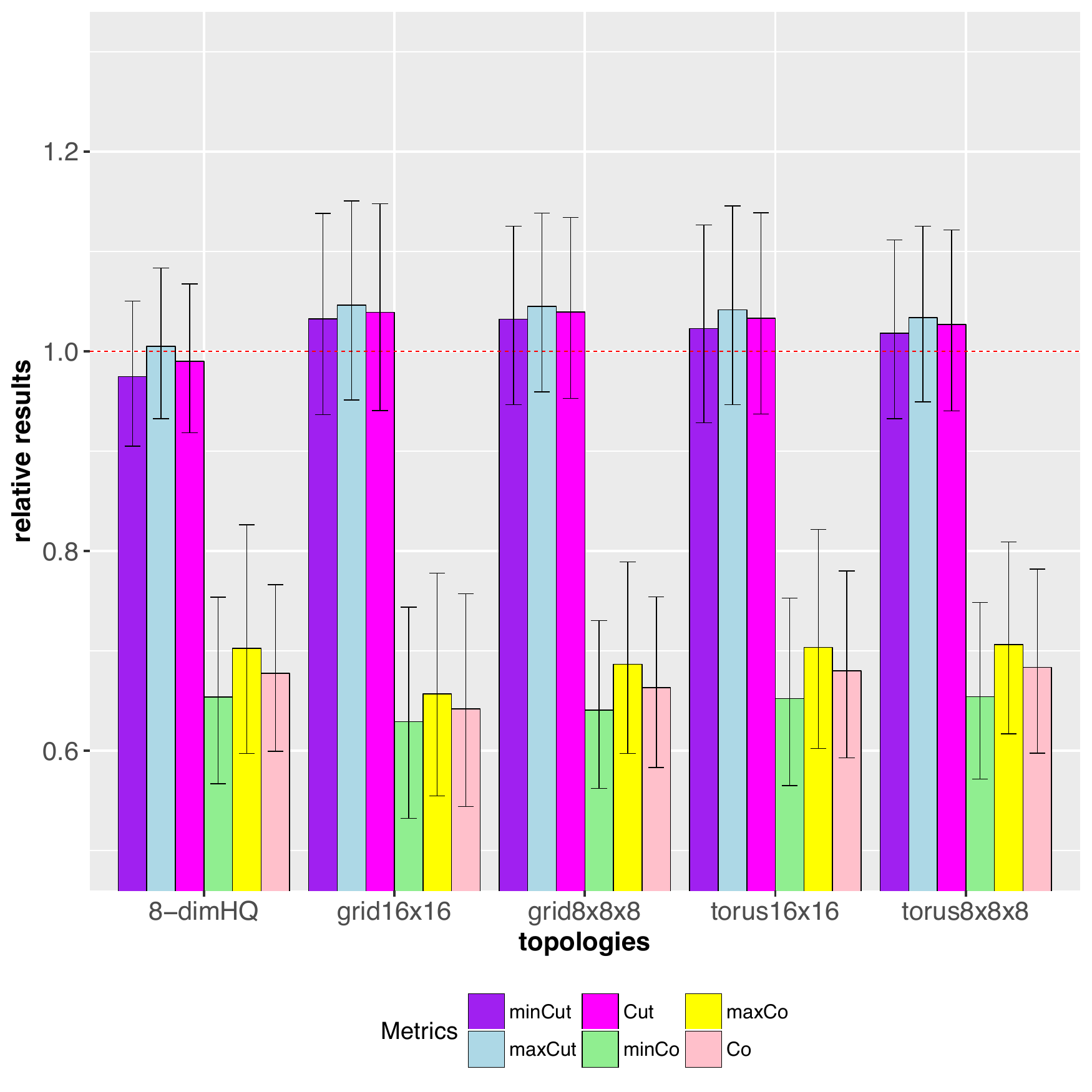}
\label{c1}
}
\subfloat[]{
\centering
\includegraphics[width=0.45\textwidth]{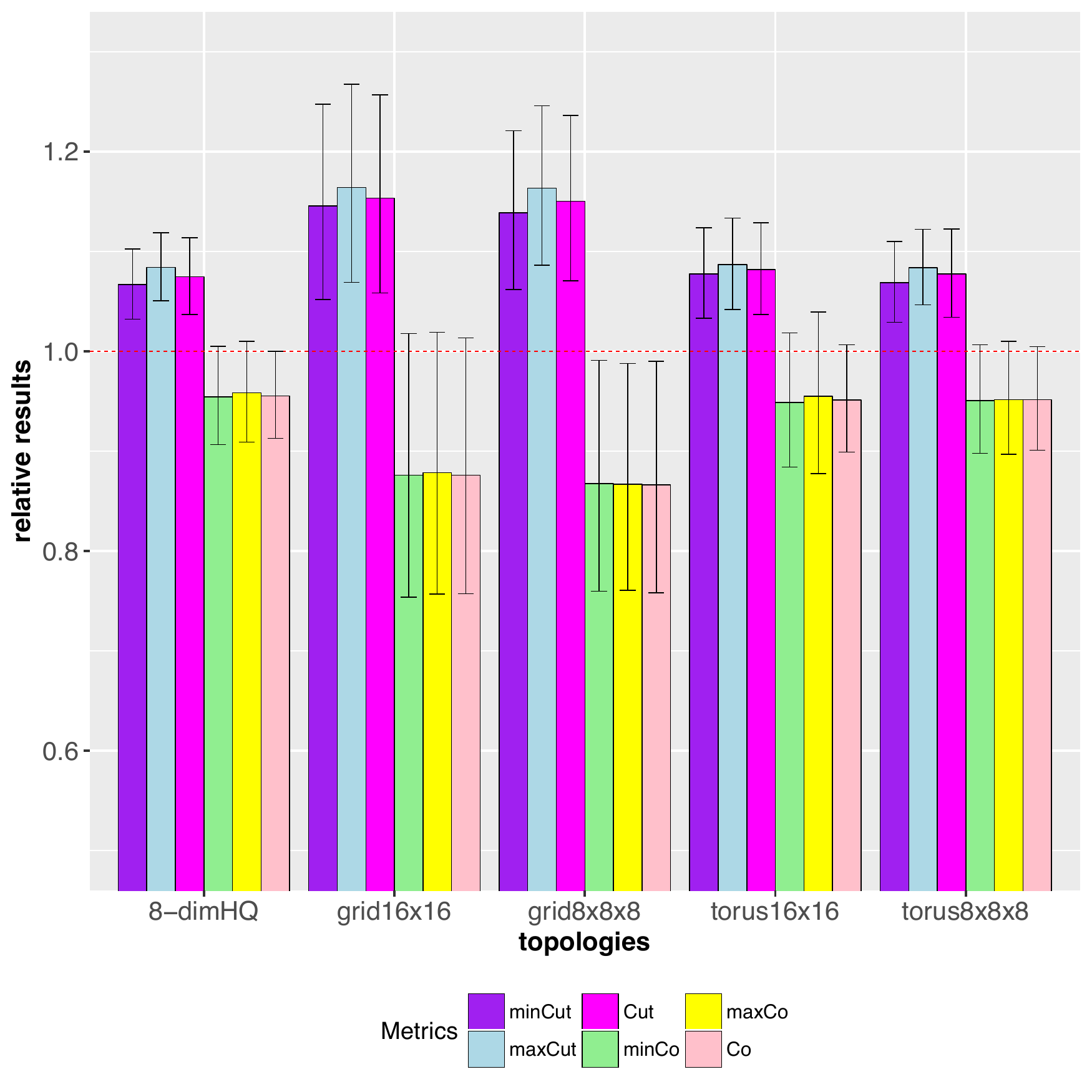}
\label{c2}
}

\subfloat[]{
\centering
\includegraphics[width=0.45\textwidth]{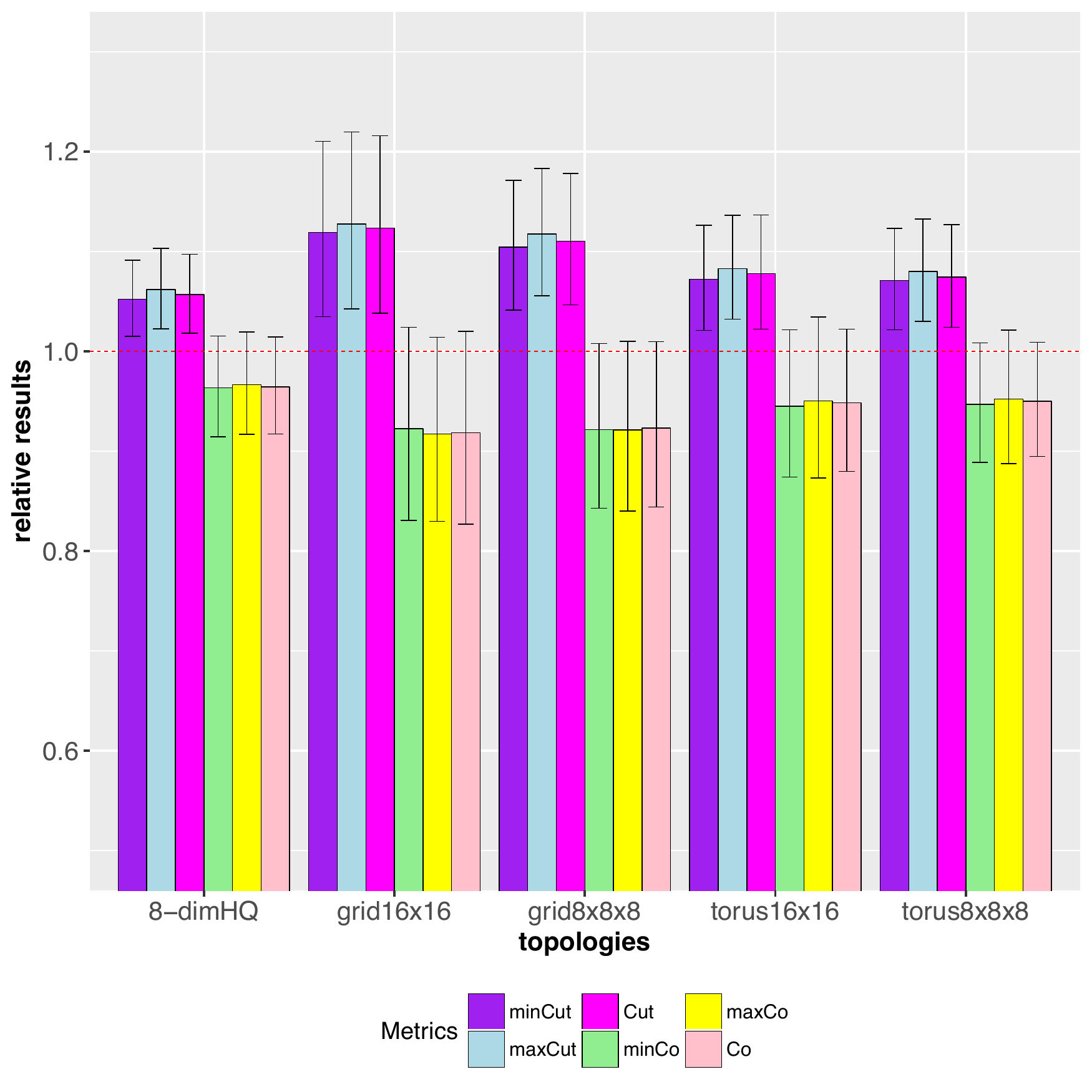}
\label{c3}
}
\subfloat[]{
\centering
\includegraphics[width=0.45\textwidth]{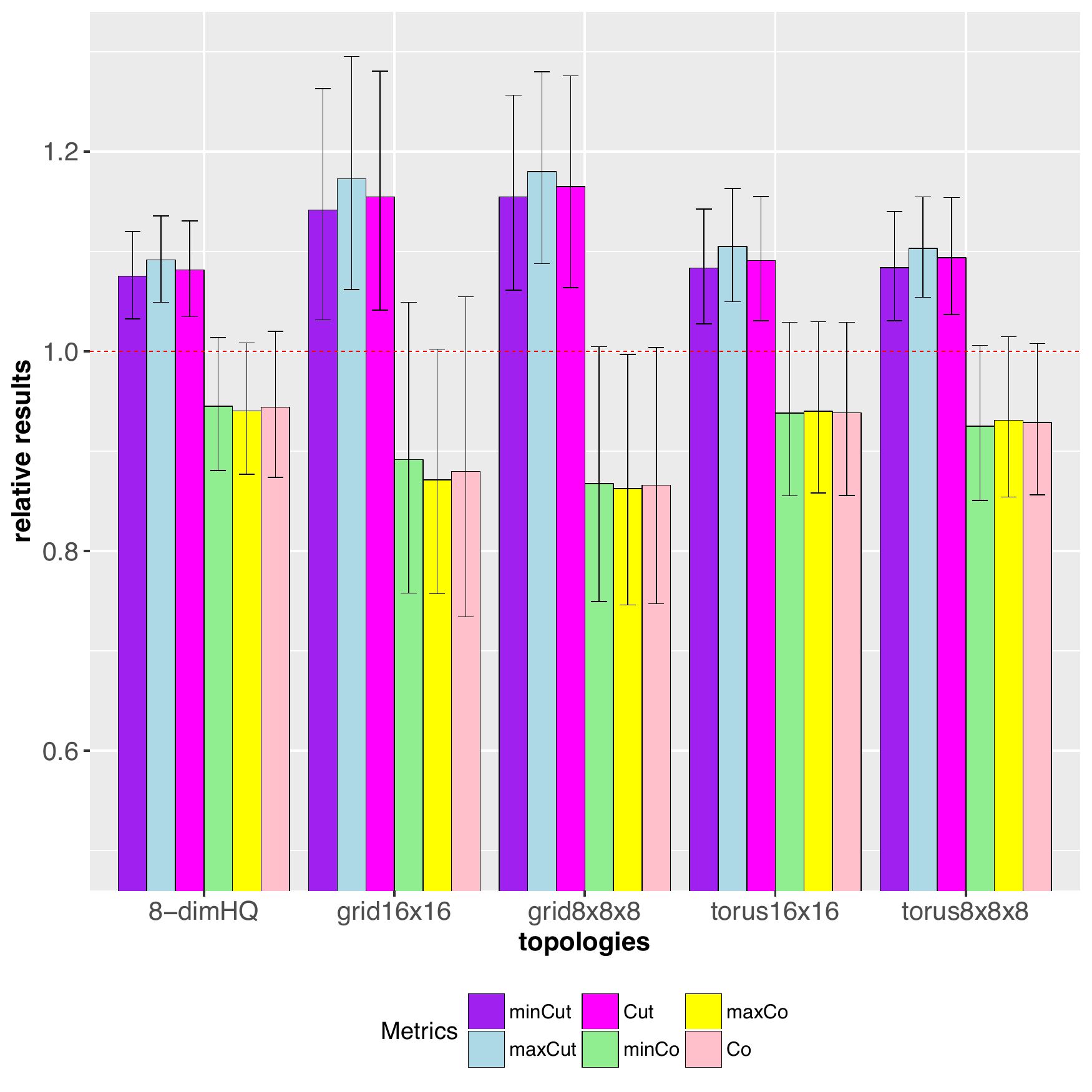}
\label{c4}
}

\caption{Quality results ($Co$ and $Cut$) for experimental case (a) $\cone$ (initial mapping with \scotch),
(b) $\ctwo$ (initial mapping with \initial), (c) $\cthree$ ((initial mapping with $\greedyallcM$)), and
(d) $\cfour$ (initial mapping with $\greedymin$).
}
\end{figure*}

{
   \scriptsize
\begin{table*}[!b]
\centering

\caption{{\small Running time results of each experimental case. For $\cone$ results are relative to \scotch's mapping time, while for $\ctwo$, $\cthree$, $\cfour$, results are relative to partitioning time with \kahip (original values in Appendix~\ref{sec:app-tables}, Table~\ref{tab:absRunTimesKahip})}}

\begin{tabular}{|l||l|l|l||l|l|l||l|l|l||l|l|l||}
\hline
& \multicolumn{3}{c||}{\scotch ($\cone$)}
& \multicolumn{3}{c||}{\initial ($\ctwo$)}
& \multicolumn{3}{c||}{$\greedyallcM$ ($\cthree$)}
& \multicolumn{3}{c||}{$\greedymin$ ($\cfour$)}

\\ \cline{2-13} 
& $qT^{gm}_{min}$ & $qT^{gm}_{mean}$ & $qT^{gm}_{max}$ & $qT^{gm}_{min}$ & $qT^{gm}_{mean}$ & $qT^{gm}_{max}$ & $qT^{gm}_{min}$ & $qT^{gm}_{mean}$ & $qT^{gm}_{max}$ & $qT^{gm}_{min}$ & $qT^{gm}_{mean}$ & $qT^{gm}_{max}$\\ \hline
$16 \times 16$ grid & 30.2780 & 29.8388 & 31.8387 & 0.95310 & 1.00480 & 1.05286       & 0.97916 & 1.01791 & 1.05075        & 0.95448 & 1.00681 & 1.05500     \\ \hline
$8 \times 8 \times 8$ grid & 18.0226 & 18.0484 & 19.5701 & 0.47495 & 0.49364 & 0.51333      & 0.47525 & 0.49427 & 0.51606    & 0.48712 & 0.50654 & 0.52698     \\ \hline
$16 \times 16$ torus & 21.1373 & 21.2507 & 22.5322 & 0.61627 & 0.64089 & 0.66334            & 0.61743 & 0.63765 & 0.66042    & 0.64834 & 0.66524 & 0.68270    \\ \hline
$8 \times 8 \times 8$ torus & 13.8136 & 14.0924 & 14.3866 & 0.33254 & 0.34167 & 0.35008     & 0.32952 & 0.33885 & 0.34855    & 0.33412 & 0.34493 & 0.35744    \\ \hline
$8$-dim HQ & 11.2948 & 11.4237 & 11.5842 & 0.36821 & 0.37977 & 0.38916                    & 0.36631 & 0.37246 & 0.38005    & 0.37254 & 0.38093 & 0.39196     \\ \hline
\end{tabular}
\label{tab:extra_social_initial_1}
\end{table*}
}

\begin{itemize}

\item When looking at the running times for the experimental cases $\ctwo$ to $\cfour$(in Table~\ref{tab:extra_social_initial_1}), we see that
  the running time results of \mswap
  are on the same order of magnitude as partitioning
  with \kahip; more precisely, \mswap is on average 42\% faster.
  Thus, inserting \mswap into the partitioning/mapping pipeline of \kahip would 
  not increase the overall running time significantly.
  
  The comparison in case $\cone$ needs to be different. Here the initial mapping is produced by
  \scotch's mapping routine (using partitioning internally), so that the relative timing results of \mswap 
  are divided by \scotch's mapping time. \scotch is much faster (on average $19$x), but its solution quality
  is also not good (see $Co$ metric in Figure~\ref{c1}). In informal experiments, we observed that only ten hierarchies (parameter $N_h$) are sufficient for \mswap
  to improve the communication costs significantly compared to \scotch -- with a much lower running time penalty than the one reported here. 
  Finally, recall that parallelization could reduce the running time of \mswap.
  
  
\item Processor graphs do have an influence on running times.
  The processor graphs, from top to bottom in Table~\ref{tab:extra_social_initial_1}, have
  30, 21, 32, 24 and 8 convex cuts, respectively. Thus, if we keep
  grids and tori apart, we can say that the time quotients increase
  with the number of convex cuts. Recall that the number of
  convex cuts equals the length of the labels of $V_p$. Moreover, the
  length of the extension of $V_p$'s labels to those of $V_a$ depends
  also on the processor graph, because a higher number of PEs
  (number of blocks) yields fewer elements per block. However, this
  influence on the length of the labels is small (increase of $1$ in
  case of $\vert V_p \vert = 256$ compared to $\vert V_p \vert =
  512$). Thus it is basically the length of $V_a$'s labels that
  determines the time quotients. For the experimental cases $\ctwo$ to $\cfour$, this observation is in line with the
  fact that \kahip takes longer for higher values of $\vert V_p
    \vert$ (see Table~\ref{tab:absRunTimesKahip} in Appendix~\ref{sec:app-tables}).


  \item \mswap successfully reduces communication costs in a range from
    6\% to 34\% over the different experimental cases (see $minCo, Co$ and $maxCo$ values in Figure~\ref{subsec:results}).
    It does so at the expense of the edge cut metric with an average increase between 2\% to 11\% depending on the experimental case.
    Note that for case $\cone$ the edge cut increase is minimum (Figure~\ref{c1}). Moreover, for cases $\ctwo$ to $\cfour$ this increase
    is not surprising due to the
  different objectives of the graph partitioner (\kahip) and \mswap. On grids
  and tori, the reduction of communication cost, as measured by
  $\cc(\cdot)$, is respectively 18\% and 13\% (on average, over all experimental cases).

  The better the connectivity
  of $G_p$, the harder it gets to improve $\cc(\cdot)$
  (results are poorest on the hypercube). (Note that
  $q_{min}$ values can be larger than $q_{mean}$ and $q_{max}$ values
  due to the evaluation process described in
  Section~\ref{subsec:setup}.)

  We observed before~\cite{Glantz2015c} that $\greedyallcM$
  performs better on tori than on grids; this is probably due to
  $\greedyallcM$ ``invading'' the communication graph \emph{and} the
  processor graph. The resulting problem is that it may paint itself into a
  corner of the processor graph (if it has corners, like a
  grid). Thus, it is not surprising that for $\ctwo$ the improvement \wrt
  $\cc(\cdot)$ obtained by \mswap is
  greater for grids than for tori. Likewise, we observe that \mswap is able to decrease
  the communication costs significantly for $\cone$ (even more than in the other cases).
  Apparently, the generic nature of \scotch's mapping approach leaves room for such an improvement.
\end{itemize}


%
%
\section{Conclusions}
\label{sec:conclusions}
We have presented a new method, \mswap, to enhance mappings of
computational tasks to PEs. \mswap can be applied whenever the
processor graph $G_p$ is a partial cube. Exploiting this property, we supply
the vertices of the application graph with labels that encode the current mapping and facilitate a straightforward assessment of any gains/losses of local search moves.
By doing so, we are able to improve initial mappings using a multi-hierarchical search method.




Permuting the entries of the vertex labels in the application graph gives
rise to a plethora of very diverse hierarchies. These
  hierarchies do not reflect the connectivity of the application
  graph $G_a$, but correspond to recursive bipartitions of $G_a$, which, in turn, are extensions of ``natural'' recursive bipartitions of $G_p$. 
  This property of \mswap suggests to
  use \mswap as a complementary tool to
  enhance state-of-the-art methods for partitioning and mapping.
  
In our experiments we were able to improve state-of-the-art mappings of complex networks
to different architectures by about $6\%$ to $34\%$ in terms of $\cc$.
  More precisely, for grids we obtained, on average, an improvement of 18\% and for tori an improvement of
  13\% over the communication costs of the initial mappings.

The novelty of \mswap consists in the way it harnesses the fact
  that many processor graphs are partial cubes: the local search
  method itself is standard and simple. We assume that further
  improvements over state-of the art mappings can be achieved by
  replacing the simple local search by a more sophisticated method.

\section*{Acknowledgments}
This work is partially supported by German Research Foundation (DFG) grant ME 3619/2-1.
Large parts of this work were carried out while H.M. was affiliated with Karls\-ruhe
Institute of Technology.


\bibliographystyle{ACM-Reference-Format}
\bibliography{roland,paper2,refs-parco}

 \clearpage 
\pagenumbering{gobble}
\appendix

\section{Appendix}
\subsection{Additional experimental results}
\label{sec:app-tables}

\begin{table}[H]
\caption{Running times in seconds for \kahip to partition the complex
  networks in Table~\ref{tab:complex} into $\vert V_p \vert = 256$ and
  $\vert V_p \vert = 512$ parts, respectively. These partitions are
  used to construct the starting solutions for the mapping algorithms for cases $\ctwo$ to $\cfour$. 
}
\begin{center}
\begin{tabular}{| l | r | r |}
\hline
    {\bf Name} & $\vert V_p \vert = 256$ & $\vert V_p \vert = 512$\\\hline \hline
        PGPgiantcompo  & 1.457 &    2.297    \\\hline
          as-22july06 & 11.179 &   13.559    \\\hline
           as-skitter & 1439.316  &  2557.827    \\\hline
     citationCiteseer  & 217.951  &  367.716    \\\hline
    coAuthorsCiteseer  & 58.120  &   69.162    \\\hline
        coAuthorsDBLP  & 157.871 &  233.000    \\\hline
     coPapersCiteseer  & 780.491 &  841.656    \\\hline
         coPapersDBLP  & 1517.283  & 2377.680    \\\hline
          email-EuAll  & 22.919 &   17.459    \\\hline
 loc-brightkite\_edges  & 113.720 &  155.384    \\\hline
    loc-gowalla\_edges  & 461.583 & 1174.742    \\\hline
       p2p-Gnutella04  & 16.377  &   17.400    \\\hline
     soc-Slashdot0902  & 887.896  & 1671.585    \\\hline
           web-Google  & 128.843 &  130.986   \\\hline
           wiki-Talk   & 1657.273 & 4044.640   \\\hline\hline
{\bf Arithmetic mean} & 498.152  & 911.673      \\\hline
{\bf Geometric mean}  & 142.714  & 204.697       \\\hline
\end{tabular}
\end{center}
\label{tab:absRunTimesKahip}
\end{table}

\clearpage 
\pagenumbering{gobble}

\end{document}